\documentclass[aps,prb,superscriptaddress,longbibliography, twocolumn, floatfix]{revtex4-2}
\usepackage{bm}
\usepackage{lmodern}

\usepackage[utf8]{inputenc}
\usepackage{amsmath}
\usepackage{siunitx}
\usepackage{amssymb}
\usepackage{graphicx}
\usepackage{epstopdf}
\usepackage{xcolor}
\usepackage{enumerate}
\usepackage{ulem}
\usepackage[english]{babel}
\usepackage{braket}
\usepackage{natbib}
\usepackage{svg}
\usepackage{float}
\usepackage{stmaryrd}
\usepackage[T1]{fontenc}

\usepackage{hyperref}
\usepackage{lipsum}
\usepackage[title]{appendix}

\begin{document}


\title{A vertical gate-defined double quantum dot in a strained germanium double quantum well}

\author{Hanifa Tidjani}
\altaffiliation{These authors contributed equally}
\affiliation{QuTech and Kavli Institute of Nanoscience, Delft University of Technology, PO Box 5046, 2600 GA Delft, The Netherlands}
\author{Alberto Tosato}
\altaffiliation{These authors contributed equally}
\affiliation{QuTech and Kavli Institute of Nanoscience, Delft University of Technology, PO Box 5046, 2600 GA Delft, The Netherlands}
\author{Alexander Ivlev}
\affiliation{QuTech and Kavli Institute of Nanoscience, Delft University of Technology, PO Box 5046, 2600 GA Delft, The Netherlands}
\author{Corentin Déprez}
\affiliation{QuTech and Kavli Institute of Nanoscience, Delft University of Technology, PO Box 5046, 2600 GA Delft, The Netherlands}
\author{Stefan Oosterhout}
\affiliation{QuTech and Netherlands Organisation for Applied Scientific Research (TNO), Delft, The Netherlands}
\author{Lucas Stehouwer}
\affiliation{QuTech and Kavli Institute of Nanoscience, Delft University of Technology, PO Box 5046, 2600 GA Delft, The Netherlands}
\author{Amir Sammak}
\affiliation{QuTech and Netherlands Organisation for Applied Scientific Research (TNO), Delft, The Netherlands}

\author{Giordano Scappucci}
\affiliation{QuTech and Kavli Institute of Nanoscience, Delft University of Technology, PO Box 5046, 2600 GA Delft, The Netherlands}
\author{Menno Veldhorst}
\affiliation{QuTech and Kavli Institute of Nanoscience, Delft University of Technology, PO Box 5046, 2600 GA Delft, The Netherlands}

\date{\today}

\begin{abstract}
Gate-defined quantum dots in silicon-germanium heterostructures have become a compelling platform for quantum computation and simulation. Thus far, developments have been limited to quantum dots defined in a single plane. Here, we propose to advance beyond planar systems by exploiting heterostructures  with multiple quantum wells. We demonstrate the operation of a gate-defined vertical double quantum dot in a strained germanium double quantum well. In quantum transport measurements we observe stability diagrams corresponding to a double quantum dot system. We analyze the capacitive coupling to the nearby gates and find two quantum dots accumulated under the central plunger gate. We extract the position and estimated size, from which we conclude that the double quantum dots are vertically stacked in the two quantum wells. We discuss challenges and opportunities  and outline potential applications in quantum computing and quantum simulation.

\end{abstract}

\maketitle

\section*{Introduction}

Semiconductor heterostructures composed of silicon and germanium have become the leading material platform for building quantum dot qubits \cite{Philips2022, Hendrickx2021, Stano2021ReviewOP}. Developments in their fabrication and operation have enabled demonstrations of high-fidelity single and two-qubit logic, multi-qubit operation, and rudimentary quantum error correction \cite{Yoneda_2018_NatNano_COherence99, Lawrie_2021_arxiv_SimultaneousDriving, Xue2022, Noiri_2022_Nature_ErrorTheshold, Hendrickx2021, Philips2022, Takeda2022, VanRiggelenPhase}. Efforts in scaling quantum dots have led to the operation of a crossbar array comprising 16 quantum dots \cite{Borsoi2022}. Furthermore, long-range quantum links may enable to interconnect modules of quantum dot arrays \cite{Vandersypen_2017_nature_HotDenseCoherent, Wang_2023_AdvancedMat_Jellybean, Borjans_2020_Nature_PhotonMediatedLink, Yoneda_2021_NatureComm_CoherentTransport}.
These developments in gate-defined quantum dots have been restricted to quantum dots defined in a single plane, however, the versatile nature of silicon-germanium heterostructures allows for further exploration. In particular, structures with multiple quantum wells can be grown, and double quantum wells of germanium \cite{Tosato2022} and silicon \cite{Laroche2015} have been realized. An open question is thus whether multi-layer heterostructures can become a relevant platform for quantum information. Here, we motivate potential applications and experimentally explore quantum dots in stacked quantum wells.

\begin{figure*}[ht!]
    \centering
    \includegraphics{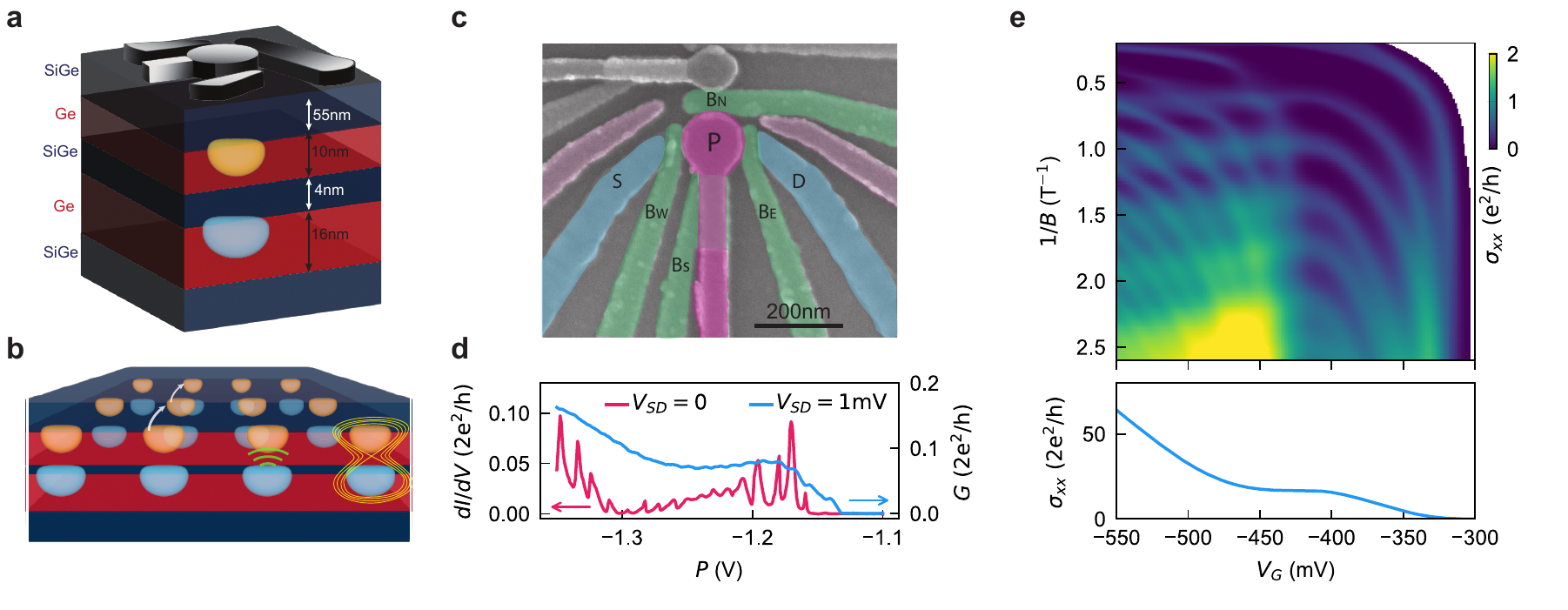}
    \caption{\textbf{Gate defined vertical double quantum dot in a bilayer heterostructure. a} Schematic of the heterostructure and gate stack. \textbf{b} Vision of a larger bilayer device, with different use-cases depicted, such as shuttling (white), sensing (green) and vertical 2-qubit gates (yellow). \textbf{c} False coloured SEM image of a similar device to the one used in this work. Quantum dots are defined under the plunger gate \textit{P} (pink) and measured in transport using the ohmic contacts source (S) and drain (D) (blue). The coupling between the quantum dot and ohmics is tuned by $B_E$ and $B_W$ (green). The potential landscape is further shaped by the gates $B_N$ and $B_S$  (green). The experiments presented in this work are performed on a section of a larger device (see Appendix \ref{supp:Full Device}). \textbf{d}  Conductance trace as a function of the plunger gate voltage through S-D at \textit{V}\textsubscript{SD}=\SI{1}{\milli V} (blue line), and differential conductance trace at \textit{V}\textsubscript{AC} = \SI{17}{\micro V}. \textbf{e} Colour map of the conductivity $\sigma_{xx}$ as a function of gate voltage $V_G$ and the inverse magnetic field $1/B$. Dark regions correspond to filled Landau levels with vanishing $\sigma_{xx}$ and correspondingly quantized $\sigma_{xy}$. Lower Panel: Linecut at $B=0$~\SI{}{T} showing the zero field conductance trace. }
    \label{fig:layout}
  
\end{figure*}

Heterostructures with parallel quantum wells may support integration of important functionalities for spin qubit based quantum processors, as depicted in Fig.\ref{fig:layout}b. Precise control over the growth of individual layers allows the engineering of inter and intralayer properties. When charges residing in separate quantum wells are capacitively coupled, but have (almost) no tunnel coupling, charge sensors could be integrated into separate layers from the qubits they sense.  In an intermediate regime where tunnel coupling is in the order of one to a few tens of gigahertz, coherent spin shuttling between the wells could be realized. Consequently, one layer may serve as quantum link for qubits defined in the other layer, for example by offering shuttling lanes that connect remote qubits \cite{Fujita_2017_npjQuantum_CoherentShuttle, Yoneda_2021_NatureComm_CoherentTransport}. In this regime the second layer can also host dedicated ancilla qubits that aid in spin-to-charge conversion for initialization and readout. Tunnel-coupled quantum wells may also be used to develop novel qubit implementations such as vertical singlet-triplet qubits or flopping mode spin qubits \cite{Mutter_2021_PRRES_FloppingMode}. Moreover, thickness and atomic composition may be tuned to optimize g-tensors\cite{Malissa2006} and spin-orbit interactions in each quantum well \cite{Scappucci2020TheRoute}, to provide dedicated functionality.

Quantum dots in multiple quantum wells may also present new opportunities for analog quantum simulation. While planar two-dimensional quantum dot arrays may be used to simulate correlated physics such as the resonating valence bond \cite{Wang2022_RVB}, quantum dots in a double quantum well may simulate even more exotic systems. For example, exciton condensation may be induced by Coulomb interactions in a regime where in one layer the quantum dot occupancy is tuned almost empty (electron layer) and the occupancy of the quantum dots in the other layer is tuned to almost filled (hole layer), where individual parameters can be controlled and studied as opposed to quantum transport implementations \cite{Conti_2021_npjQuantum_BilayerBEC, Nandi2012}. Quantum dots in multi-layer structures comprised of three or more quantum wells could also be envisioned. The confinement of quantum dots in three layers potentially supports artificial superconductivity. Attractive Coulomb interaction in quantum dot systems has been observed in planar systems \cite{Hamo2016,Hong_2018_PRB_AttractiveCoulomb} and integration of such interactions into a three quantum well system may provide a route toward tunable and controllable superconducting condensation.

These motivations warrant the study of quantum dots defined in multilayer heterostructures for quantum information. However, there are also many challenges in the fabrication, design, and operation that need to be understood and overcome. In particular, how to address individual quantum dots and tune their inter and intra layer coupling needs further exploration. We take a first step and demonstrate a vertical double quantum dot in a strained germanium double quantum well heterostructure. Through quantum transport measurements, we obtain charge stability diagrams consistent with a double quantum dot. We characterise the capacitive interaction of the quantum dots to the surrounding gates and determine their location. The size of one of the quantum dots is estimated through bias-spectroscopy. Together, these findings point to the formation of a double quantum dot vertically aligned under the same plunger gate.

\section{Results}
\begin{figure*}[htbp]
    \centering
    \includegraphics[width=\textwidth]{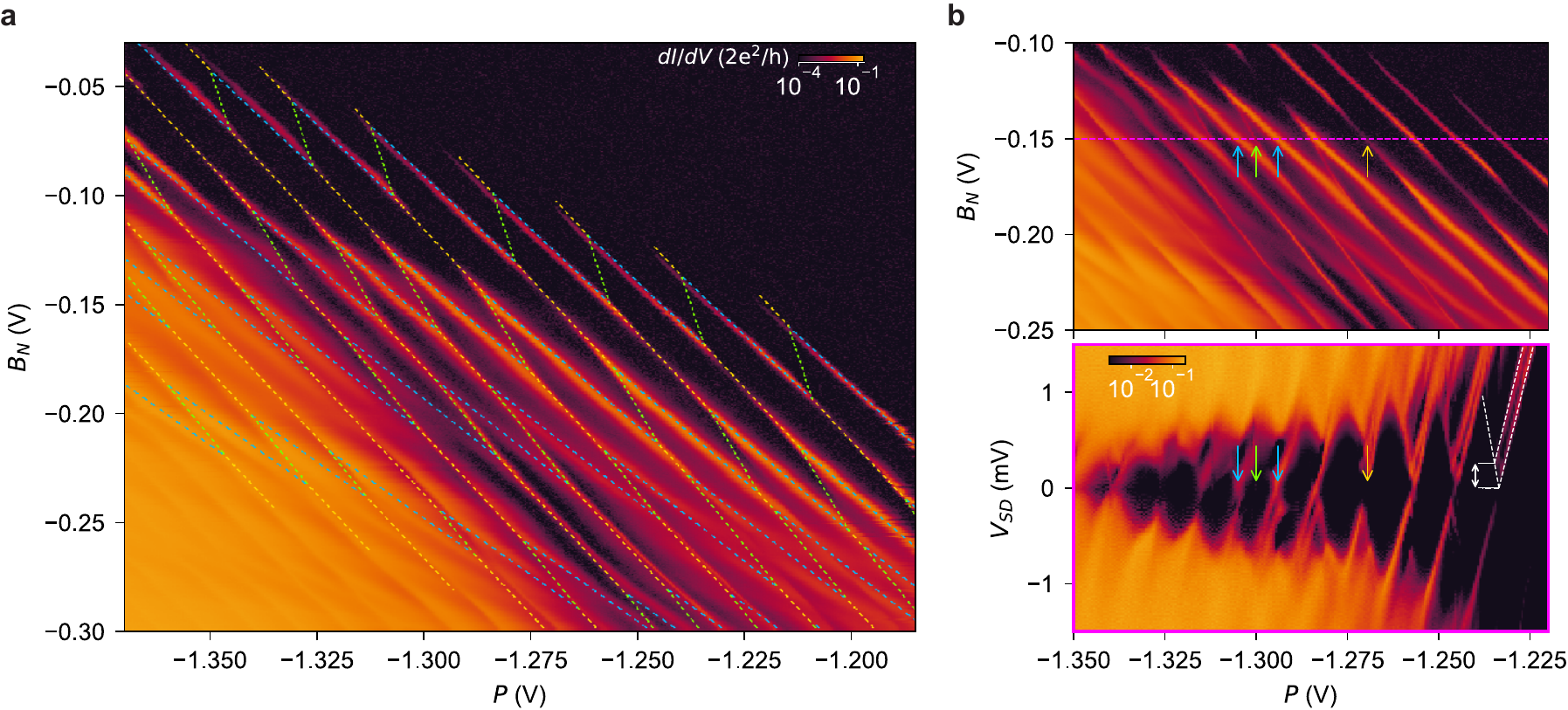}
    \caption{\textbf{Charge stability diagram of the vertical double quantum dot. a} By sweeping the gates $P$ vs $B_N$, a honeycomb pattern emerges as the gates induce transport resonances. The dashed coloured lines overlain on the data correspond to the  electrochemical simulation of the double-quantum dot system. \textbf{b} A zoom-in of the data presented in panel a with coloured arrows corresponding to different transitions (the scale bar remains the same as in \textbf{a}). \textbf{c} Bias spectroscopy across a line-cut of panel b top (magenta) at $B_N=-0.15$V, the coloured arrows correspond to the transitions highlighted in the top panel. Measurement of the orbital energy for the first Coulomb diamond is indicated by the white lines and is extracted to be approximately \SI{260}{\micro e V}. }

    \label{fig:ChargeStability}
\end{figure*}

An undoped and compressively-strained Ge/SiGe double quantum well heterostructure is epitaxially grown on a \SI{100}{mm} Si(100) substrate. A \SI{55}{nm} thick $\mathrm{Si_{0.2}Ge_{0.8}}$ spacer separates the bilayer system from the gate stack. The top and bottom quantum wells in the bilayer are \SI{10}{nm} and \SI{16}{nm} thick respectively, and are separated by a \SI{4}{nm} $\mathrm{Si_{0.2}Ge_{0.8}}$ barrier (Fig.~\ref{fig:layout}a). We perform magnetotransport characterization of a Hall-bar shaped heterostructure field effect transistor to infer the energy spectrum of the hole bilayer. The conductance map in Fig. \ref{fig:layout}e reveals the emergence of two sets of Landau levels typical of such a bilayer system~\cite{Tosato2022, Hamilton1995TransitionHeterostructures}. At $V_G\approx \SI{-320}{mV}$ the longitudinal conductivity $\sigma_{xx}$ shows the first set of quantized Landau levels, corresponding to the subband localized in the bottom well. At $V_G\approx$ \SI{-400}{mV} the conductance curve at zero-magnetic-field (Fig. \ref{fig:layout}e bottom panel) deviates from a linear increase and flattens out as the sub-band localized in the top well starts being populated. This originates from the electric field screening caused by the accumulation of charge carriers in the top well, while its density is still below the percolation threshold and transport is only available through the bottom well~\cite{Tosato2022}. For more negative voltages, the carriers in the top well start contributing to transport and conductance increases.

We then fabricate gate defined quantum dots (see methods) to probe the properties of electrostatically confined holes in this bilayer system. A 3D schematic depicting the heterostructure and gate stack, and a scanning electron microscopy (SEM) image of the device are respectively shown in Fig.~\ref{fig:layout}a,c. The central plunger gate \textit{P} is negatively biased to accumulate holes beneath it, while the barrier gates \textit{$B_W$} and \textit{B\textsubscript{E}} are used primarily to tune the tunnel barrier to the ohmic contacts (S and D). The gates \textit{B\textsubscript{N}} and \textit{B\textsubscript{S}} further shape the potential landscape without significantly affecting the tunnel barrier to the ohmics. We measure the transport through the device with DC and standard low-frequency lock-in techniques (see Methods). Similarly to the 2D transport measurement, at high source-drain DC bias (\textit{V\textsubscript{SD}}= \SI{1}{\milli \volt}, Fig.~\ref{fig:layout}d blue line), the conductance trace starts to increase as $P$ is lowered, and flattens out at P = \SI{-1.2}{V}, before increasing again as a second transport channel opens. The differential conductance ($dI/dV$) at zero DC bias (pink line), reveals the emergence of Coulomb peaks, and the formation of quantum dots. Interestingly, we observe that the Coulomb oscillation amplitude significantly decreases around the plateau. These observations are consistent with the Hall bar experiments, where first the bottom quantum well is being populated, followed by the population of the upper quantum well, which initially does not contribute to transport.

\begin{figure*}[ht!]
    \centering
    \includegraphics[width=.9\linewidth]{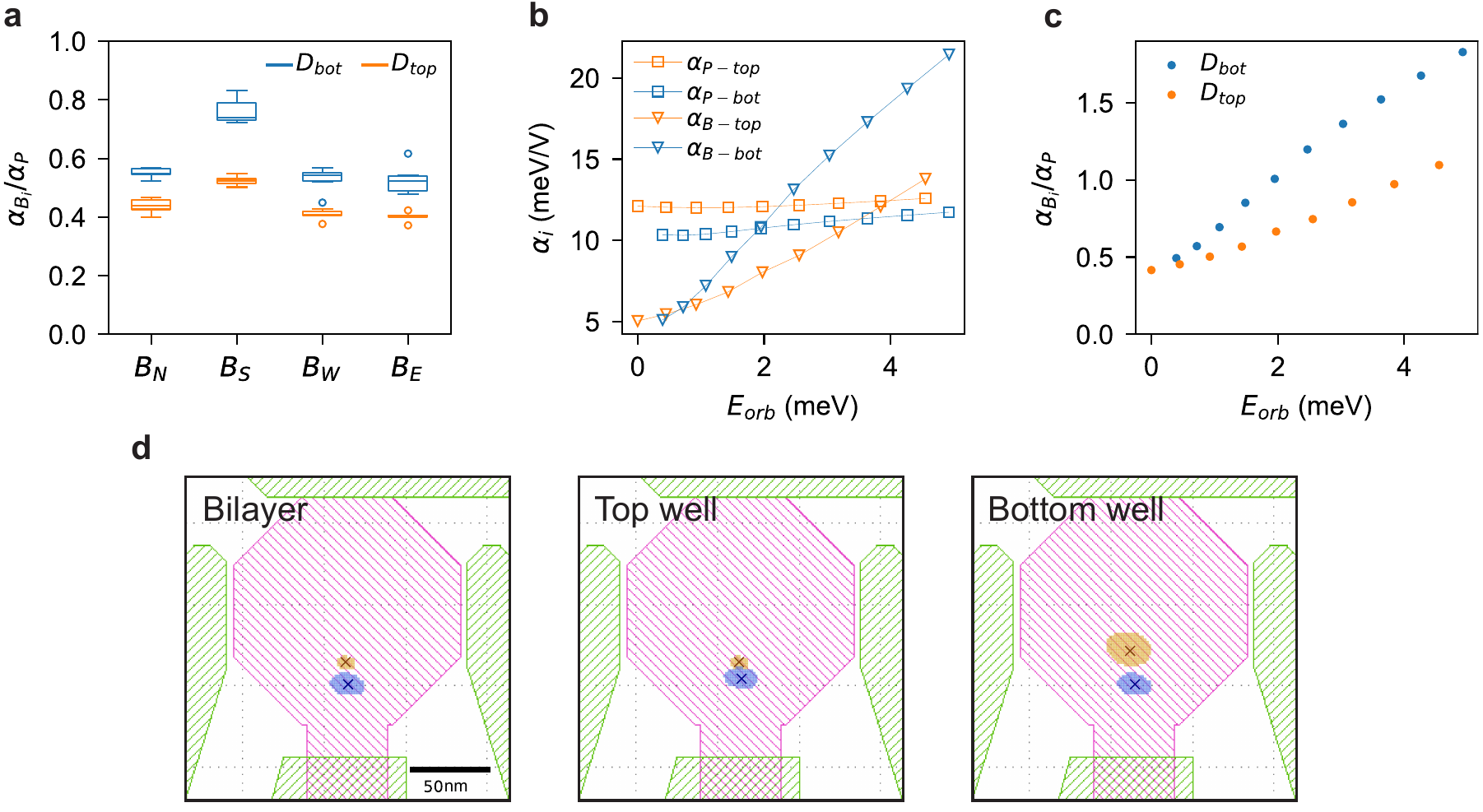}
    \caption{\textbf{Measured and simulated relative lever arms of the surrounding gates to the quantum dots, and the triangulation of their centre points. a} Boxplot of the relative lever arm ($\alpha_{B_i,D_i}/\alpha_{P,D_i}$) of each barrier gate $B_i$ to the plunger gate $P$ for the top and bottom quantum dot ($D_i$). These couplings are extracted from the slopes of different reservoir transition in the charge stability diagrams as a function of $B_i$ and $P$ (see Appendix \ref{app:AppendixB_OtherRegimes}). \textbf{b} Simulated absolute lever arm for the plunger gate $P$ and barrier gate B to the top and bottom quantum dot calculated from the Schr\"odinger-Poisson simulation described in Appendix \ref{appendix:SchrodingerPoisson}.\textbf{c} Simulated relative lever arm $\alpha_{B_i,D_i}/\alpha_{P,D_i}$ for the first 20 orbital states, plotted against the orbital energy ($E_{orb}$). Data are extracted from the Schr\"odinger-Poisson simulation described in Appendix \ref{appendix:SchrodingerPoisson}.  \textbf{d} Triangulation of the position of the quantum dots based on the coupling in \textbf{a} and the capacitive simulation (Appendix \ref{appendix:FEM}). The cross indicates the centre point of each quantum dot, and the coloured area represents the 1$\sigma$ standard deviation of this value. In the `Bilayer' panel the quantum dots are simulated in separate wells, with the orange quantum dot and blue quantum dot placed in the top and bottom well respectively. In the `Top Well' and `Bottom Well' panel the quantum dots are simulated in the same layer.}
    \label{fig:CouplingsAndPositions}

\end{figure*}

To further investigate the nature of the quantum dots in this bilayer system we map the charge stability diagram as a function of the gate voltages applied on $B_N$ and $P$ (Fig.~\ref{fig:ChargeStability}). A distinct honeycomb pattern emerges, indicating the presence of a double quantum dot system~\cite{VanDerWiel2003}. Unlike a typical double-quantum dot charge stability diagram~\cite{VanDerWiel2003} where each quantum dot predominantly responds to its own dedicated plunger gate, all transition lines in this diagram have a similar slope, indicating that both quantum dots have similar capacitance response to the gates. This observation is consistent with a vertically stacked quantum dot system, where both quantum dots are expected to have a similar capacitance to the surrounding gates. We further map the charge stability diagram as a function of $B_i$ ($i= W, S, E$) and $P$ for the remaining barrier gates (see Appendix Fig.\ref{fig:AppendixB:SimOnDifferentCS}) and similarly, the transition lines indicate a comparable capacitive coupling of the surrounding gates to the two quantum dots labelled as $D_{bot}$ and $D_{top}$. The absence of transition lines strongly coupled to the gates $B_i$ confirms that both quantum dots are centrally located under the plunger gate $P$, rather than originating from  spurious quantum dots positioned under one of the $B_i$ gates.

To distinguish the inter-dot transitions from the reservoir transitions in the charge stability diagram of Fig.~\ref{fig:ChargeStability}a, we measure the differential conductance as a function of \textit{V}\textsubscript{SD} and $P$. The bottom panel of Fig.~\ref{fig:ChargeStability}b shows a set of Coulomb diamonds taken in the regime where $B_N=\SI{-0.15}{V}$ (magenta line-cut shown in the top panel). The transition lines which fall on this line-cut correspond to \textit{V}\textsubscript{SD}=0 in the lower panel, and in both panels are indicated by the four coloured arrows. The transition lines indicated by the blue and orange arrows correspond to the edges of the Coulomb diamonds, and thus to dot-reservoir transitions. In contrast, the transition line indicated by the green arrow lies in the middle of a Coulomb diamond, and a conductance peak is barely visible. This is the expected behaviour for an inter-dot transition, where transport is via co-tunneling processes and which results in a weak conductance signal only when the two quantum dots are in resonance. Using these four transition lines as a starting point, we assign to each line in the charge stability diagrams the respective transition type.



From the slopes of the reservoir transitions in each charge stability diagram we extract the capacitive couplings for the two quantum dots to the barrier gates $B_i$ relative to the plunger gate $P$ defined as $\alpha_{B_i,D}/\alpha_{P,D}$, where $
\alpha_{B_i,D}$ is the lever arm of the barrier gate $B_i$ to quantum dot $D$. Fig. \ref{fig:CouplingsAndPositions}a shows a box-plot of the relative coupling for the two quantum dots for each barrier gate $B_i$, with the colours matching the two sets of reservoir transitions identified in the charge stability diagrams of Fig.~\ref{fig:AppendixB:SimOnDifferentCS}. We observe that the coupling of the two quantum dots to all barrier gates is lower than their coupling to the plunger gate. This is consistent with both quantum dots being located under the plunger gate. Furthermore the relative coupling of quantum dot $D_{bot}$ is larger than the coupling of $D_{top}$ for all barrier gates which is evidence of the two quantum dots being vertically stacked under the plunger gate. If the two quantum dots were not vertically stacked and instead lay in the same plane, one quantum dot would exhibit a larger relative coupling to one or two adjacent barrier gates, while the other quantum dot would show a larger relative coupling to the remaining barrier gates.

We further support this interpretation by estimating the position of both quantum dots using an electrostatic finite element method (FEM) simulation in Ansys Q3D\cite{Ansys}, in which the heterostructure, gate-layers and insulating layers are included (details can be found in Appendix \ref{appendix:FEM}). The quantum dots are simulated individually in the different layers, to reflect their possible locations (Fig.\ref{fig:CouplingsAndPositions}d). Each quantum dot is modelled as a metallic disk as thick as the quantum well it is located in. The radius and position of this simulated metallic quantum dot is varied to analyze its effect on the capacitance. The geometric capacitance between this quantum dot and the gates is determined, and assumed to be directly proportional to the lever arm $\alpha_{G,D}$. By comparing the simulated capacitance with the extracted relative coupling $\alpha_{B_i,D}/\alpha_{P,D}$, the position of a single quantum dot within either Ge layer is triangulated. The positions of the quantum dots best matching the experimental data are relatively close to each other, under the plunger gate, and both positioned towards $B_S$ (Fig.~\ref{fig:CouplingsAndPositions}d) as indicated by the cross. The centre-centre distance between the quantum dots has an upper bound of \SI{30}{nm} with $1\sigma$ standard deviation, independent of which layer the simulation is performed on, as indicated by the coloured area. The standard deviation is based on the spread of the relative couplings that is extracted from each charge-stability diagram. The prohibitively close proximity of the center of these two quantum dots suggests that two quantum dots are located in the two different wells.

Furhermore, from the orbital energy $E_{orb} =  \SI{260}{\micro e \volt}$ extracted in \ref{fig:ChargeStability}b (white lines) for the quantum dot corresponding to the blue transitions, we  estimate the dot size. Assumimg a harmonic in-plane potential $V_{xy} = \frac{1}{2}m^{*}\omega^2(x^2+y^2)$, where $\hbar\omega = E_{orb}$ and $m^*=0.055m_e$, this gives a quantum dot $d$ diameter of about $d=\sqrt{\hbar/(\omega m^{*})} = \SI{137}{nm}$, comparable to the plunger gate size of 150nm. Based on the size approximation of this quantum dot and their mutual proximity, we conclude that the quantum dots cannot coexist in a single layer without coalescing. Overall, the interpretation of the measured relative capacitive couplings, along with the results from the FEM simulation and the estimates of the quantum dot size from the Coulomb diamonds, provide strong arguments for the quantum dots being vertically stacked under the plunger gate.

To gain further insight into this vertical double-quantum dot system we perform a 2D Schr\"odinger-Poisson simulation and present the results in Fig.~\ref{fig:CouplingsAndPositions}~b,c. Fig.~\ref{fig:CouplingsAndPositions}c shows the relative capacitive coupling of the quantum dots formed in the top quantum dot ($D_{top}$) and bottom quantum dot ($D_{bot}$) for different orbitals as a function of orbital energy. These relative couplings are calculated from the absolute capacitive couplings of $D_{bot}$ and $D_{top}$ to the barrier gate $B$ and the plunger gate $P$ shown in panel Fig.~\ref{fig:CouplingsAndPositions}b. While the absolute coupling of the plunger gate to the top and bottom quantum dot ($\alpha_{P-top}$ and $\alpha_{P-bot}$) remains approximately constant with increasing orbital number, the barrier gate lever arm to both quantum dots varies significantly with the orbital number. This is because an increasing orbital number corresponds to an increase of the wavefunction radius. As a result the distance to the barrier gate gets smaller and the coupling to the barrier gate increases. Panel b shows that a relatively larger lever arm of the barrier gates is expected when the quantum dot is located in the bottom quantum well. We find this for all the simulated orbitals, suggesting that in Fig.~\ref{fig:CouplingsAndPositions}a $D_{top}$ corresponds to a quantum dot located in the upper quantum well and $D_{bot}$ to a quantum dot located in the bottom quantum well.

To confirm the position of the quantum dots suggested by the Schr\"odinger-Poisson simulation, we look at the inter-dot transitions in the charge stability diagrams of Appendix Fig.~\ref{fig:AppendixB:SimOnDifferentCS}. In all diagrams the inter-dot line strongly couples to the plunger gate, with a hole being transported from $D_{bot}$ (blue transition lines) to the $D_{top}$ (orange transitions lines) at more negative voltages. This provides evidence that $D_{top}$ is localized in the top well, with a hole being attracted from the bottom quantum dot towards the plunger gate on top.

\section{Discussion and outlook}

In this work we demonstrate that a vertical double quantum dot can be formed and controlled in double quantum well heterostructure. A single gate can be used to simultaneously populate quantum dots in two quantum wells whilst the charge occupation can be tuned using one of the surrounding gates. This provides prospects for quantum dot arrays in multiple quantum wells. Integration of charge sensors may allow to tune to the single-hole regime. The separation of the quantum wells may be used as a coarse parameter to tune the interlayer coupling between the dots, while the observation of different lever arms corresponding to the wells suggest that gate voltages may be used for further tuning. Establishing quantum dot arrays beyond planar arrays may provide new means for quantum computation and simulation with quantum dots. 


\section{Acknowledgements}
We acknowledge useful discussions with members of the Veldhorst, Scappucci and Vandersypen group. We thank S. G. J. Philips and S. De Snoo for the development of the software stack used to aquire data in this experiment. 

\section{Data Availability}

The raw data and analysis supporting the findings of this study are openly available in a Zenodo repository: https://zenodo.org/record/7962368. 

\section{Funding}
We acknowledge support through an NWO ENW grant and an ERC Starting Grant. 

\section{Competing Interests}
We declare there are no competing interests. 

\section{Methods}
The device is fabricated on a Si\textsubscript{$x$}Ge\textsubscript{$1-x$}/Ge/Si\textsubscript{$x$}Ge\textsubscript{$1-x$}/Ge/Si\textsubscript{$x$}Ge\textsubscript{$1-x$} heterostructure, where $x = 0.2$, grown by reduced pressure chemical vapour deposition. The virtual substrate upon which the heterostructure is grown consists of a silicon substrate, upon which there is a $1.6$\textmu m relaxed Ge layer; a 1 \textmu m graded Si\textsubscript{$x$}Ge\textsubscript{$1-x$} layer, with final Ge composition of x = 0.2.  On top of the SiGe virtual substrate, the bilayer system comprises of a \SI{16}{nm} thick bottom Ge quantum well, a \SI{4}{nm} thick SiGe barrier, a \SI{10}{nm} thick top Ge quantum well, and a final \SI{55}{nm} thick SiGe spacer. At the top of the stack a sacrificial Si cap is grown to provide a native SiOx oxide layer. We define ohmic contacts using electron beam lithography and buffered oxide etch of the silicon cap layer. We then evaporate a \SI{30}{nm} Platinum layer and using a 10 minute rapid thermal anneal at $400^\circ $ C, contact the quantum well. The ohmic layer is isolated using a \SI{7}{nm} layer of Al\textsubscript{2}O\textsubscript{3} grown by atomic layer deposition.   Electrostatic gates used to define the quantum dots are defined in two layers, (3/\SI{17}{nm} and 3/\SI{37}{nm} of Ti/Pd.) and are separated by a \SI{5}{nm} layer of Al\textsubscript{2}O\textsubscript{3}. 

Devices are screened at \SI{4}{K} using standard dipstick measurements. Experiments reported in this paper are carried out in a Bluefors LD400 dilution refrigerator with a base temperature of \SI{10}{mK}. The electrical properties are investigated through two terminal AC and DC measurements. There is a tunable DC voltage component \textit{V}\textsubscript{SD} used for biasing the device, and an oscillating AC voltage is applied when taking measurements with a lock-in amplifier. The differential conductance $dI/dV_{\textrm{SD}}$ is measured using standard lock-in techniques with a typical frequency of \SI{70}{Hz}, and an amplitude of \SI{17}{\micro V}.

\onecolumngrid\
\begin{appendices}
\setcounter{table}{0}
\setcounter{figure}{0}

\setcounter{section}{0}
\section{Full Device}

 \label{supp:Full Device}
\begin{figure*}[hbtp!]
    \centering
    \includegraphics{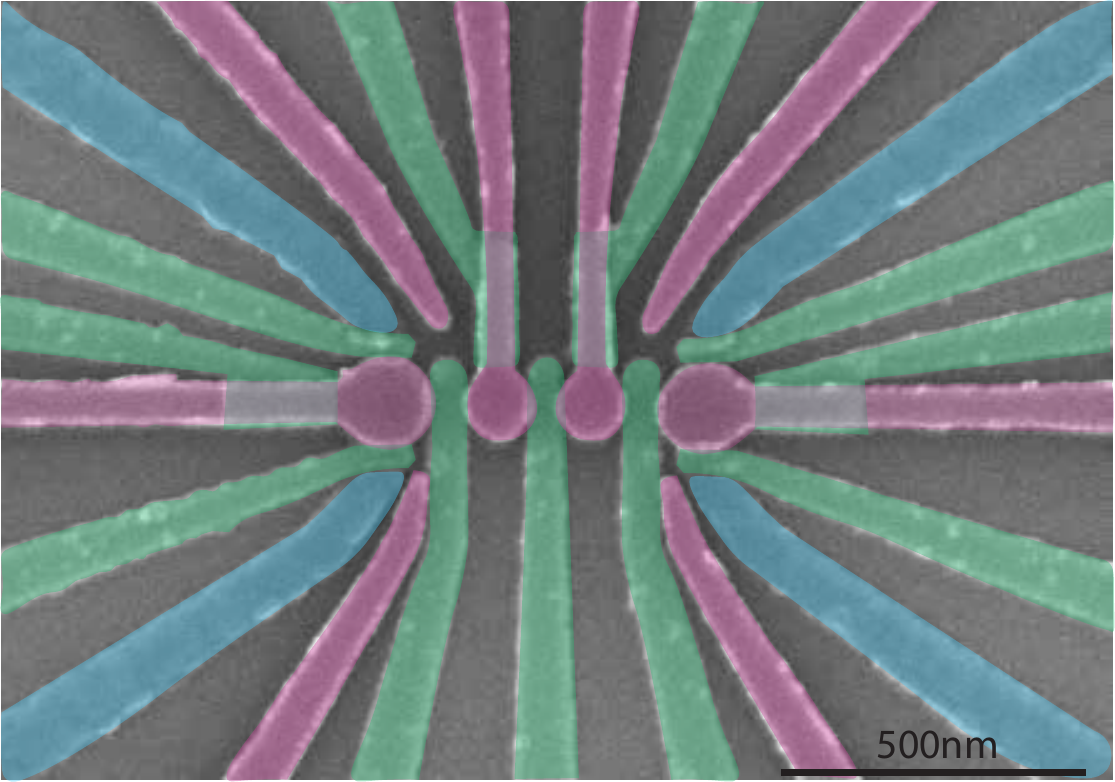}
    \caption{\textbf{A false coloured SEM of a device similar to the one used in this experiment}. Experimentally we focus on the left part of the device, and study quantum transport through the left quantum dot (purple) coupled to the ohmics (blue) and controlled via barrier gates (green). }
    \label{fig:full_dev}
\end{figure*}

\section{Further discussion on charge stability diagrams}
\label{app:AppendixB_OtherRegimes}
\begin{figure*}[htbp!]
    \centering
    \includegraphics[width=0.75\textwidth]{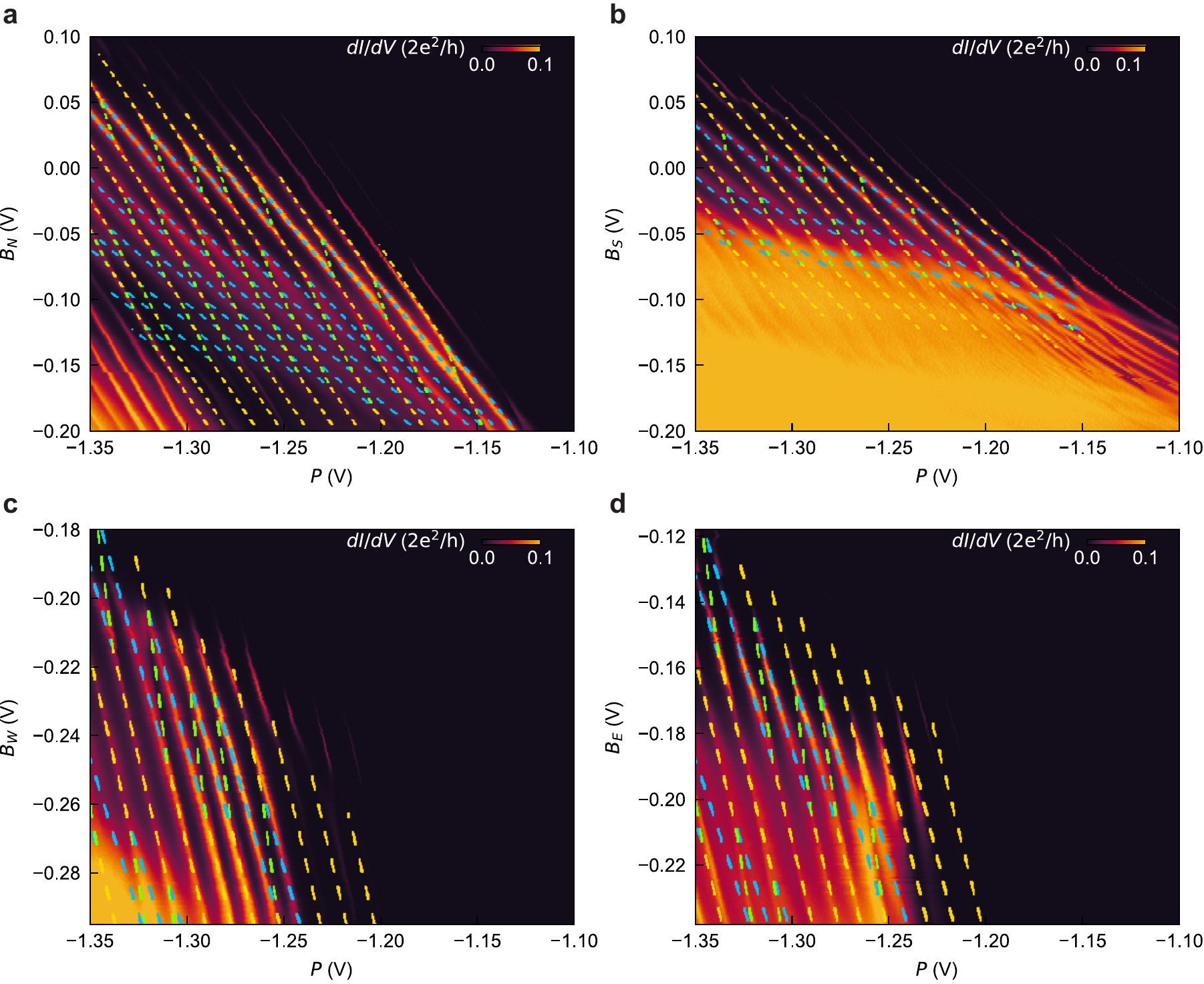}
    \caption{\textbf{Charge stability diagrams as a function of the Plunger gate $P$, and barrier gates $B_N$, $B_S$, $B_W$, $B_E$, with electrochemical simulation overlain.} Comparison between transport data and the double quantum dot simulation defined by equation \ref{eq:chem_potentail}. Subfigures \textbf{a-d} respectively give the conduction through the system as function of the voltages on the plunger gate $P$ against the voltage respectively applied on $B_N,B_S,B_W$ and $B_E$. These plots share a common barrier gate reference at $\mathbf{V_{ref}}=P = -1.217\text{V}, B_N=0\text{V},B_S=0\text{V},B_W=-0.237\text{V},B_E=-0.178\text{V}]$. The correspondence between the model and data is particularly convincing for the $B_N$ and $B_S$ data sets, which could sweep a broader region as they influence the quantum dot coupling to the reservoirs less. The simulation has been limited to the double-quantum dot region. Data without the simulation is found in Fig. \ref{fig:AppendixB:DifferentCS}. The code used to reproduce these simulations can be found on Zenodo via the link provided in the main manuscript. }
    \label{fig:AppendixB:SimOnDifferentCS}
\end{figure*}

In order to triangulate the position of both quantum dots as described in the main part of the paper, charge stability diagrams (CSD) were taken by the variation of the plunger gate $P$ against all surrounding barrier gates (Fig. \ref{fig:AppendixB:SimOnDifferentCS}). These diagrams are all taken in a similar gate-region, enabling comparisons of the relative lever arms between the gates and the quantum dots, as is shown in the main. To show that these are compatible with a double quantum dot, the four CSDs are overlain with a shared electrochemical simulation of a double quantum dot system. The details of the simulation are found in appendix \ref{app:ElectrochemicalSimulation}. 
While the $P-B_N$ CSD is similar to the one in the main text, the quantum dot transitions are less pronounced in this regime. This may be attributed to a change in the effective tunnel-coupling in this different region. 
\\
The CSDs presented here as well as in the main article also have voltage regions that are poorly predicted by a double quantum dot model, in particularly at more negative values for $B_N$ and $B_S$. For the negative $B_N$ region the transitions associated with bottom-well quantum dot fade out or disappear entirely. A possible reason could be the coupling between the two layers, as a significant tunnel coupling may delocalise the quantum dot and alter the transport behavior. At even more negative $B_N$ and $P$ voltages, the transitions associated with the top quantum dot strongly dominate. This is thought to correspond to the localisation of the wavefunction in the top well, similar to existing Schr\"odinger-Poisson simulation of a Hall-bar system and accompanying measurements \cite{Tosato2022}. 

In the $B_S-P$ map (Fig. \ref{fig:AppendixB:SimOnDifferentCS}b), single-well dominated transport also appears in the more negative voltage region, although a large background current is observed as well. This background current emerges as the bottom quantum dot resonances become poorly defined and closely spaced. This can be attributed to an increasing size of the second quantum dot, leading to decreasing charging energy and stronger coupling to the leads. This behaviour can be expected given the relatively large lever arm of $B_S$ to the quantum dots.

\begin{figure*}[htbp!]
    \centering
    \includegraphics[width=0.75\textwidth]{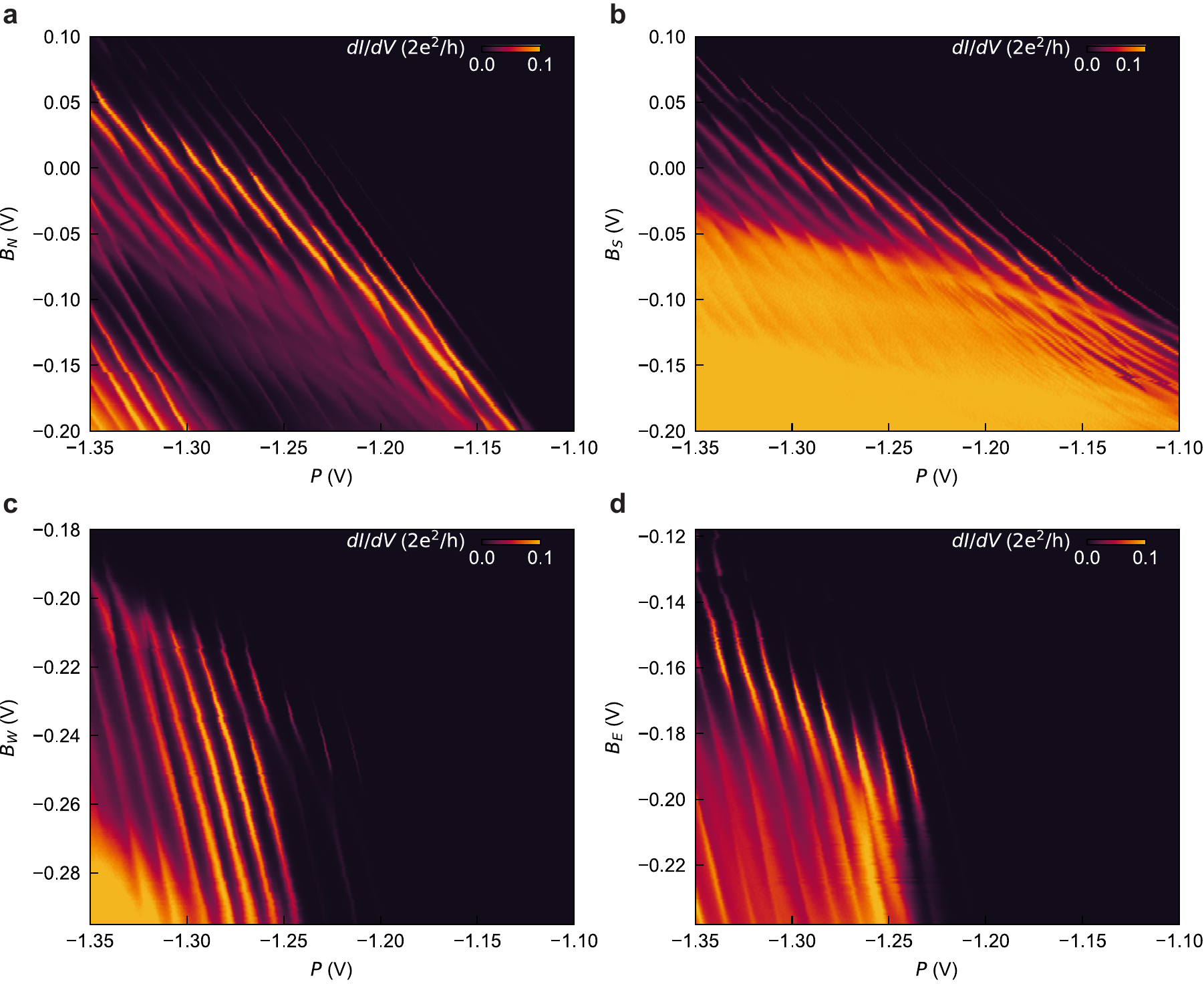}
    \caption{\textbf{Charge stability diagrams as a function of the Plunger gate $P$, and barrier gates $B_N$, $B_S$, $B_W$, $B_E$, without electrochemical simulation overlain.} }
    \label{fig:AppendixB:DifferentCS}
\end{figure*}

\section{Electrochemical Double quantum dot Simulations}
\label{app:ElectrochemicalSimulation}
The charge-stability simulation of a double-quantum dot was made using a classical electrochemical picture \cite{VanDerWiel2003}. The charge-carriers in the simulation are holes, and gates couple to both quantum dots through the occupation-dependent lever arm matrix $\alpha$. In the model used, the electrochemical potential $\mu$ of the quantum dots is given by:
\begin{align}
    \begin{split}
        \mu_1(N_1,N_2,\mathbf{V})&=\Sigma_{n_1=1}^{N_1} E_{C1}(n_1)\\
        &\quad+\Sigma_{n_2=1}^{N_2} E_{Cm}(N_1,n_2)\\
        &\quad+\alpha_1(N_1,N_2,\mathbf{V})\mathbf{V}\\
        \mu_2(N_1,N_2,\mathbf{V})&=\Sigma_{n_2=1}^{N_2} E_{C2}(n_2)\\
        &\quad+\Sigma_{n_1=1}^{N_1} E_{Cm}(n_1,N_2)\\
        &\quad+\alpha_2(N_1,N_2,\mathbf{V})\mathbf{V}
    \end{split}
    \label{eq:chem_potentail}
\end{align}
$N_{1(2)}$ denotes the occupation of the top (bottom) quantum dot, with $E_{C1(2)}(N_{1(2)})$ and $E_{Cm}(N_1,N_2)$ the respective occupation dependent charging energies of the top and bottom quantum dots, and the electrostatic inter-dot coupling energy. $\mathbf{V}$ is the vector containing the gate-voltages. The coupling between the gates and the top (bottom) quantum dot is given by two row-matrices $\alpha_{1(2)}$, with purely positive entries. As indicated in equation \ref{eq:chem_potentail}, the lever arms are generally taken to be occupation- and voltage-dependent. Since the charging energy is poorly-defined for level-dependent lever-arms, these are defined at an arbitrary fixed reference voltage $\mathbf{V_{ref}}$: 
\begin{align}
    \begin{split}
            E_{C1(2)}(N_{1(2)})
            &\equiv\mu_{1(2)}(N_{1(2)},N_{2(1)},\mathbf{V_{ref}})\\
    &\quad-\mu_{1(2)}(N_{1(2)}-1,N_{2(1)},\mathbf{V_{ref}})
    \end{split}
\end{align}
using $\mu(0,0,\mathbf{V_{ref}})=0$. The charging energies and the inter-dot capacitive energies, are fixed when reproducing the datasets of figure \ref{fig:AppendixB:SimOnDifferentCS} as these are in the same regime. This reduces the amount of free parameters, preventing overfitting the data. To reduce the free parameters further, the dependencies of the coupling matrix $\alpha$ can be simplified. $\alpha_1(V_{P})$ and $\alpha_2(N_2,V_{P})$ can simply depend on the plunger gate voltage $V_{P}$, with the latter also on the occupation of the bottom quantum dot, as this was necessary to match the data. With these assumptions, the correspondence seen in figure \ref{fig:AppendixB:SimOnDifferentCS} was achieved. These values are by no means unique, and are included in view of transparency. Moreover, they give a qualitative indication of the behaviour of the system, illustrating that the parameters are relatively constant for a fixed occupation of the bottom quantum dot. This indicates a well-behaved system, which is well explained by just two quantum dots.

The parameter that changes the most across the voltage range is the relative coupling strength to the gates, as function of the bottom quantum dot occupancy. The loading of the bottom quantum dot increases the size of its wavefunction leading to an effectively stronger coupling to the barrier gates relative to the plunger gate coupling. The large impact of the bottom quantum dot occupation compared to the top quantum dot might be explained by the relative difference in quantum dot size and overall charge occupation.

We also want to note that the sizable effect of the plunger gate voltage on the relative coupling to the south barrier gate indicates that the quantum dots move more centrally under the plunger gate. This suggests that the quantum dots are close to the south barrier gate as expected in our main analysis. Further discrepancies between the data and the model are attributed to the limited parameters of the model, like the neglected dependence on the top quantum dot occupation and the fact that tunnel-coupling is disregarded. Moreover as mentioned in appendix \ref{app:AppendixB_OtherRegimes}, some regions are poorly described by a double quantum dot. Given the correspondence to this relatively simple model we are convinced that we observe a double quantum dot model across multiple occupations. This justifies the analysis of the charge stability diagram in light of such a system. The code and the parameters used to perform the simulation can be found in full on Zenodo using the link provided in the main text. There one also finds the parameters to fit the charge stability diagram in figure \ref{fig:ChargeStability} of the main text. 

\section{Schr\"odinger-Poisson simulation}
\label{appendix:SchrodingerPoisson}
\begin{figure*}[htbp!]
    \centering
    \includegraphics{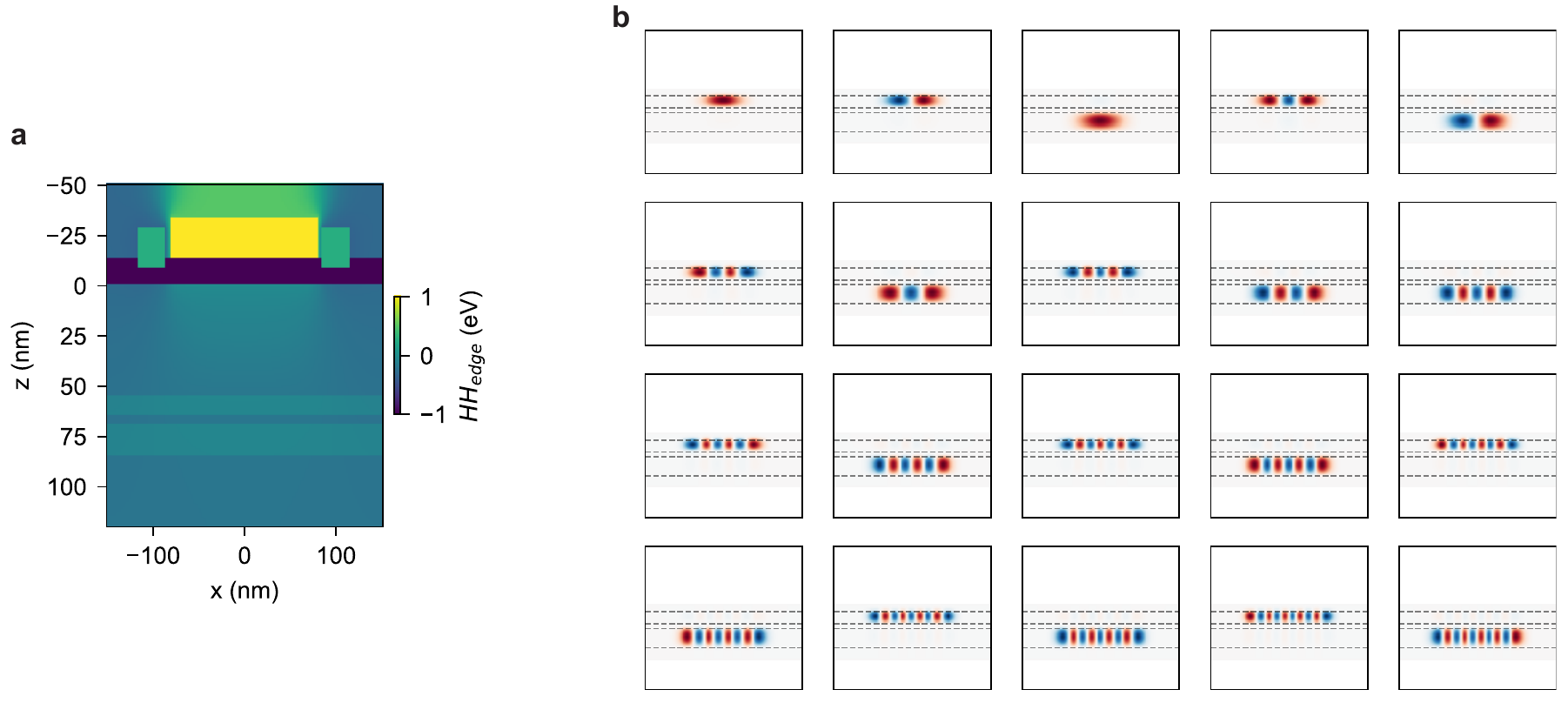}
    \caption{\textbf{Results of the 2D Schr\"odinger-Poisson simulations.} \textbf{a} Heavy hole band edge as a function of x and z coordinates for the bilayer heterostructure which also comprises the gate stack for $z<0$, where $z$ is the growth direction. \textbf{b} Wavefunction amplitude for the first 20 energy levels ordered from left to right, top to bottom.}
    \label{fig:Schr}
\end{figure*}
We perform 2D Schr\"odinger-Poisson simulation with Nextnano~\cite{nextnano}. Figure~\ref{fig:Schr}a shows the heavy-hole (HH) band edge for the heterostructure ($z>0$) where the top edges of the quantum wells are positioned at z=\SI{55}{nm} and z=\SI{69}{nm}. For $z<0$ the gate stack is visible. This comprises a plunger gate, two barrier gates and a layer of Al$_2$O$_3$. Panel \ref{fig:Schr}b shows the wavefunction amplitude for the first 20 2D orbitals for which we calculated the lever arm. The potential applied to the plunger and to the barriers is \SI{-1.15}{V} and \SI{-0.25}{V} respectively.

\section{Finite Elements Simulations of Capacitance}
\label{appendix:FEM}
\begin{figure*}[h!]
    \centering
    \includegraphics[width=\textwidth]{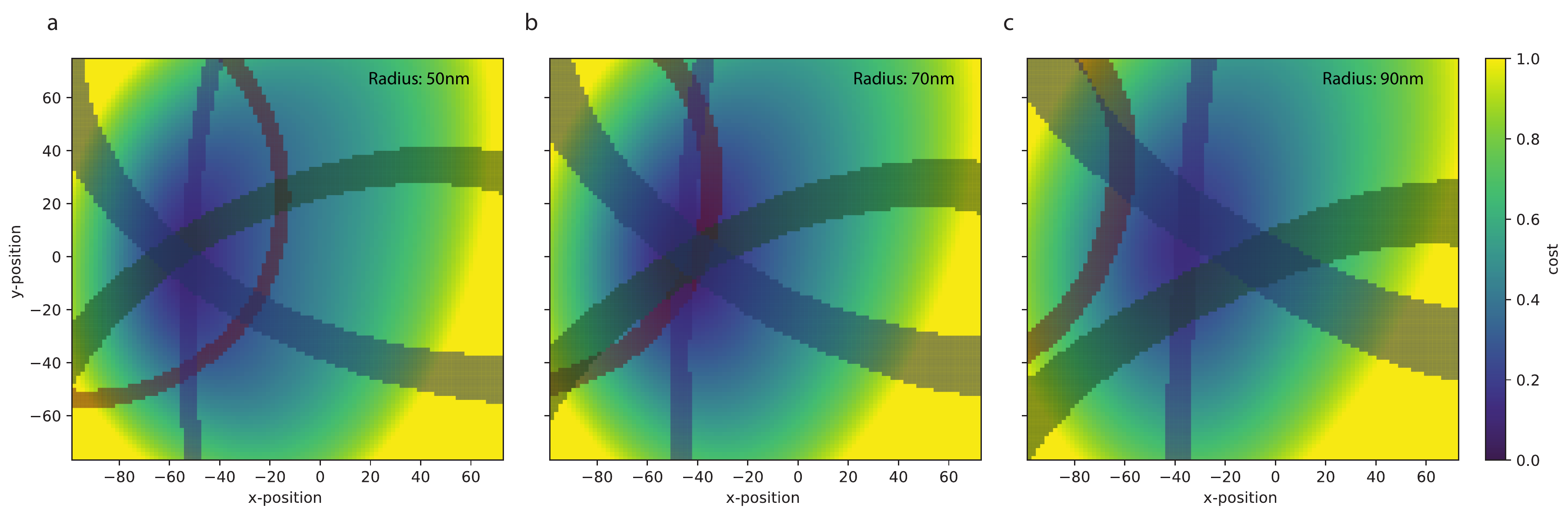}
    \caption{\textbf{Triangulation of the dot location based on the capacitive coupling to the surrounding gates.} The contour plots indicate the most likely position of the simulated quantum dot based on the with the difference in the measured and simulated relative gate-couplings. The most likely position is at the point of the lowest the cost, defined as $\sqrt{\sum_{B_i}(\Tilde{\alpha}_{B_i,\text{meas}}-\Tilde{\alpha}_{B_i,\text{sim}})^2}$, where $\Tilde{\alpha}_{B_i,\text{meas(sim)}}$ is the measured (simulated) relative lever arm $\alpha_{B_i}/\alpha_P$ and the sum is taken over the barrier gates $B_N$, $B_S$, $B_E$ and $B_W$. The origin is taken at the center of the plunger gate. Subfigures \textbf{a, b, c} respectively indicate the results for 50, 70 and 90 nm quantum dot radii.  The red, purple, green and blue contours correspond to the best positions providing simulated relative couplings within the standard deviation of the empirically extracted mean values $c_{B_N}=0.55,c_{B_S}=0.76,c_{B_W}=0.53,c_{B_E}=0.53$ (Fig.~\ref{fig:CouplingsAndPositions} of the main text). In these simulations, the quantum dot has been assumed to be in the top well. Similar results are obtained for quantum dot 1 and for the quantum dots in the bottom well. One notices that as the quantum dot-size grows, the best-position contours move away from their respective gates, as one would expect. It is also clear that the contours overlap with each other for the 70 nm quantum dot, making this the predicted radius in case quantum dot 2 would be placed in the top well.}
    \label{fig:AppendixD:Positionquantum dot1_topwell}
\end{figure*}

To get an indication of the position of the quantum dots based on their relative coupling to the gates, a finite element simulation with Ansys Q3D \cite{Ansys} is performed. Using this simulation the geometric capacitance of each quantum dot is extracted. These simulated geometric capacitances are converted to relative lever arms by dividing the absolute quantum dot-barrier gate capacitance by the quantum dot-plunger gate capacitance. The simulated relative lever arms are then compared to those extracted from the charge-stability diagrams. The position that best matches the measured relative gate couplings is determined, as exemplified in Fig. \ref{fig:AppendixD:Positionquantum dot1_topwell} for quantum dot 1. This is done for the quantum dot being positioned in either the top or the bottom well. The heterostructure has been modelled up to \SI{500}{nm} in-plane around the plunger gate, and up to a depth of \SI{125}{nm} below the bottom quantum well. The dielectric material permittivities used are found in table \ref{tab:AppendixD:Permittivity}. The metallic gates have been simulated as uniform perfect conductors.

\begin{table}[htbp!]
\caption{\textbf{Relative electric Permittivity of the dielectric materials.} These are the values used in the Ansys simulation. For the properties of Si\textsubscript{0.2}Ge\textsubscript{0.8} a linear approximation based on the atomic concentration is taken. }
\begin{tabular}{|l|l|}
\hline
Material   & $\epsilon_r$ \\ \hline
Al\textsubscript{2}O\textsubscript{3}\cite{Groner_ThinSolidFilms_2002_AlOxProperties}       & 5.9            \\ \hline
Ge\cite{Dunlap_PhysRev_1953_SiGeDielectricConstants}         & 15.8            \\ \hline
Si\textsubscript{0.2}Ge\textsubscript{0.8} \cite{Dunlap_PhysRev_1953_SiGeDielectricConstants}  & 15.0            \\ \hline
SiO\textsubscript{2}\cite{Sze_1981_PropertiesSiO2}  & 3.9            \\ \hline
\end{tabular}
\label{tab:AppendixD:Permittivity}
\end{table}
In each instance of the simulation a circular, perfectly conducting, uniform disk was placed in either quantum well to emulate the quantum dot. This allows to estimate the geometric capacitance between the quantum dot and the surrounding gates. The radius of the simulated quantum dots and the best performing size is taken for the analysis of each quantum dot. We note that the wavefunction density of the quantum dot is not taken into account here, and non-circular shapes are not investigated. 

Future improvements of this method would consider the electrostatic potential arising from the gates, the strain of the system as well as the effects of the disorder. Moreover, the interplay between multiple quantum dots spread across the wells can be taken into account in the future, as we stress that currently just a single quantum dot is simulated at a time. A self-consistent Schr\"odinger-Poisson based approach could make this possible.

\begin{figure*}[h]
    \centering
    \includegraphics[width=\textwidth]{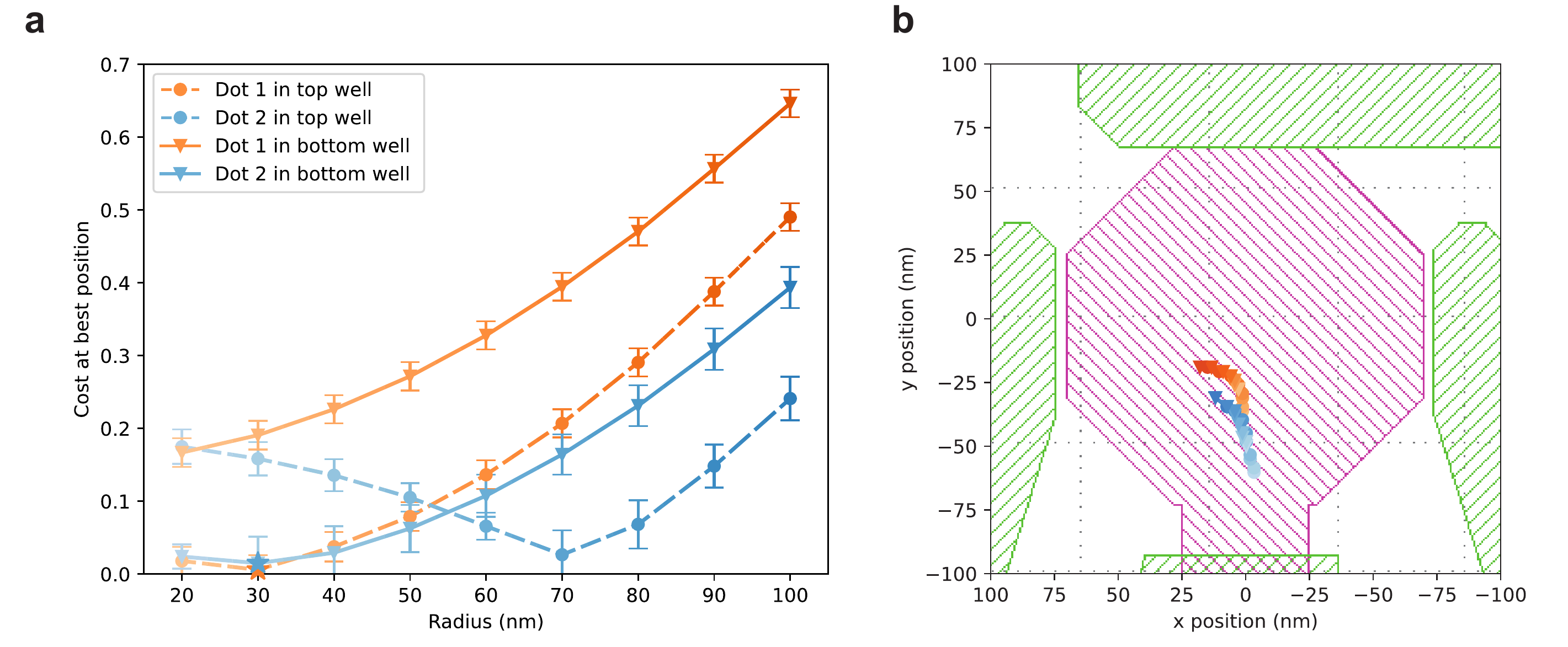}
    \caption{\textbf{Most likely quantum dot radius and position in capacitive FEM simulation.} \textbf{a} The sum of the squared difference of the coupling between the simulation and measurements, for different simulated quantum dot radii. The measured coupling to both quantum dots have been compared, simulated in both quantum wells. quantum dot 1 (2) corresponds to the orange (blue) transitions in the Fig. \ref{fig:ChargeStability} of the main text, as well as $D_{Top}$ ($D_{Bot}$) in Fig. \ref{fig:CouplingsAndPositions} of the main text. For each dot, the most likely radius has been indicated with a star. Errorbars are based on standard deviation in the slopes extracted from the raw data in \ref{fig:AppendixB:SimOnDifferentCS}. \textbf{b} The expected position for each radius of the two quantum dots in either the bottom or the top well. It is clear that for any radius the quantum dots are close to one another.}
    \label{fig:Expected_radius_comparison}
\end{figure*}

As expected, there is a particular quantum dot position and radius at which all simulated couplings agree with the measurements. These are considered to be the most probable locations and radii of the quantum dots. To compare different radii, as well as the two layers of the heterostructure, the minimum costs at the most likely location is presented in Fig.~\ref{fig:Expected_radius_comparison} for each quantum dot, radius and layer. The cost is defined as $\sqrt{\sum_{B_i}(\Tilde{\alpha}_{B_i,\text{meas}}-\Tilde{\alpha}_{B_i,\text{sim}})^2}$, where $\Tilde{\alpha}_{B_i,\text{meas(sim)}}$ is the measured (simulated) relative lever arm $\alpha_{B_i}/\alpha_P$ and the sum is taken over the barrier gates $B_N$, $B_S$, $B_E$ and $B_W$. 

Using this cost-metric, quantum dot 2 (blue) is predicted to most likely be in the bottom well and have approximately 30 nm radius, while quantum dot 1 (orange) is most likely in the top well, also with a 30 nm radius. While wells associated with the dots match our expectation of the system, the radius is lower than expected from the size of the plunger gate. However, since the exact wavefunction density is neglected in the estimation of the quantum dot-radius, it should be viewed as a comparison of the two quantum dots with each other. When we compare the quantum dots as they are simulated in the same layer, we indeed see that quantum dot 2 is simulated to be bigger as one expects from the stronger coupling to the barrier gates.

\end{appendices}

\begin{thebibliography}{32}%
\makeatletter
\providecommand \@ifxundefined [1]{%
 \@ifx{#1\undefined}
}%
\providecommand \@ifnum [1]{%
 \ifnum #1\expandafter \@firstoftwo
 \else \expandafter \@secondoftwo
 \fi
}%
\providecommand \@ifx [1]{%
 \ifx #1\expandafter \@firstoftwo
 \else \expandafter \@secondoftwo
 \fi
}%
\providecommand \natexlab [1]{#1}%
\providecommand \enquote  [1]{``#1''}%
\providecommand \bibnamefont  [1]{#1}%
\providecommand \bibfnamefont [1]{#1}%
\providecommand \citenamefont [1]{#1}%
\providecommand \href@noop [0]{\@secondoftwo}%
\providecommand \href [0]{\begingroup \@sanitize@url \@href}%
\providecommand \@href[1]{\@@startlink{#1}\@@href}%
\providecommand \@@href[1]{\endgroup#1\@@endlink}%
\providecommand \@sanitize@url [0]{\catcode `\\12\catcode `\$12\catcode
  `\&12\catcode `\#12\catcode `\^12\catcode `\_12\catcode `\%12\relax}%
\providecommand \@@startlink[1]{}%
\providecommand \@@endlink[0]{}%
\providecommand \url  [0]{\begingroup\@sanitize@url \@url }%
\providecommand \@url [1]{\endgroup\@href {#1}{\urlprefix }}%
\providecommand \urlprefix  [0]{URL }%
\providecommand \Eprint [0]{\href }%
\providecommand \doibase [0]{http://dx.doi.org/}%
\providecommand \selectlanguage [0]{\@gobble}%
\providecommand \bibinfo  [0]{\@secondoftwo}%
\providecommand \bibfield  [0]{\@secondoftwo}%
\providecommand \translation [1]{[#1]}%
\providecommand \BibitemOpen [0]{}%
\providecommand \bibitemStop [0]{}%
\providecommand \bibitemNoStop [0]{.\EOS\space}%
\providecommand \EOS [0]{\spacefactor3000\relax}%
\providecommand \BibitemShut  [1]{\csname bibitem#1\endcsname}%
\let\auto@bib@innerbib\@empty
\bibitem [{\citenamefont {Philips}\ \emph {et~al.}(2022)\citenamefont
  {Philips}, \citenamefont {Mądzik}, \citenamefont {Amitonov}, \citenamefont
  {de~Snoo}, \citenamefont {Russ}, \citenamefont {Kalhor}, \citenamefont
  {Volk}, \citenamefont {Lawrie}, \citenamefont {Brousse}, \citenamefont
  {Tryputen}, \citenamefont {Wuetz}, \citenamefont {Sammak}, \citenamefont
  {Veldhorst}, \citenamefont {Scappucci},\ and\ \citenamefont
  {Vandersypen}}]{Philips2022}%
  \BibitemOpen
  \bibfield  {author} {\bibinfo {author} {\bibfnamefont {S.~G.~J.}\
  \bibnamefont {Philips}}, \bibinfo {author} {\bibfnamefont {M.~T.}\
  \bibnamefont {Mądzik}}, \bibinfo {author} {\bibfnamefont {S.~V.}\
  \bibnamefont {Amitonov}}, \bibinfo {author} {\bibfnamefont {S.~L.}\
  \bibnamefont {de~Snoo}}, \bibinfo {author} {\bibfnamefont {M.}~\bibnamefont
  {Russ}}, \bibinfo {author} {\bibfnamefont {N.}~\bibnamefont {Kalhor}},
  \bibinfo {author} {\bibfnamefont {C.}~\bibnamefont {Volk}}, \bibinfo {author}
  {\bibfnamefont {W.~I.~L.}\ \bibnamefont {Lawrie}}, \bibinfo {author}
  {\bibfnamefont {D.}~\bibnamefont {Brousse}}, \bibinfo {author} {\bibfnamefont
  {L.}~\bibnamefont {Tryputen}}, \bibinfo {author} {\bibfnamefont {B.~P.}\
  \bibnamefont {Wuetz}}, \bibinfo {author} {\bibfnamefont {A.}~\bibnamefont
  {Sammak}}, \bibinfo {author} {\bibfnamefont {M.}~\bibnamefont {Veldhorst}},
  \bibinfo {author} {\bibfnamefont {G.}~\bibnamefont {Scappucci}}, \ and\
  \bibinfo {author} {\bibfnamefont {L.~M.~K.}\ \bibnamefont {Vandersypen}},\
  }\href {\doibase 10.1038/s41586-022-05117-x} {\bibfield  {journal} {\bibinfo
  {journal} {Nature}\ }\textbf {\bibinfo {volume} {609}},\ \bibinfo {pages}
  {919} (\bibinfo {year} {2022})}\BibitemShut {NoStop}%
\bibitem [{\citenamefont {Hendrickx}\ \emph {et~al.}(2021)\citenamefont
  {Hendrickx}, \citenamefont {Lawrie}, \citenamefont {Russ}, \citenamefont {van
  Riggelen}, \citenamefont {de~Snoo}, \citenamefont {Schouten}, \citenamefont
  {Sammak}, \citenamefont {Scappucci},\ and\ \citenamefont
  {Veldhorst}}]{Hendrickx2021}%
  \BibitemOpen
  \bibfield  {author} {\bibinfo {author} {\bibfnamefont {N.~W.}\ \bibnamefont
  {Hendrickx}}, \bibinfo {author} {\bibfnamefont {W.~I.~L.}\ \bibnamefont
  {Lawrie}}, \bibinfo {author} {\bibfnamefont {M.}~\bibnamefont {Russ}},
  \bibinfo {author} {\bibfnamefont {F.}~\bibnamefont {van Riggelen}}, \bibinfo
  {author} {\bibfnamefont {S.~L.}\ \bibnamefont {de~Snoo}}, \bibinfo {author}
  {\bibfnamefont {R.~N.}\ \bibnamefont {Schouten}}, \bibinfo {author}
  {\bibfnamefont {A.}~\bibnamefont {Sammak}}, \bibinfo {author} {\bibfnamefont
  {G.}~\bibnamefont {Scappucci}}, \ and\ \bibinfo {author} {\bibfnamefont
  {M.}~\bibnamefont {Veldhorst}},\ }\href {\doibase 10.1038/s41586-021-03332-6}
  {\bibfield  {journal} {\bibinfo  {journal} {Nature}\ }\textbf {\bibinfo
  {volume} {591}},\ \bibinfo {pages} {580} (\bibinfo {year}
  {2021})}\BibitemShut {NoStop}%
\bibitem [{\citenamefont {Stano}\ and\ \citenamefont
  {Loss}(2021)}]{Stano2021ReviewOP}%
  \BibitemOpen
  \bibfield  {author} {\bibinfo {author} {\bibfnamefont {P.}~\bibnamefont
  {Stano}}\ and\ \bibinfo {author} {\bibfnamefont {D.}~\bibnamefont {Loss}},\
  }\href@noop {} {\bibfield  {journal} {\bibinfo  {journal} {Nature Reviews
  Physics}\ }\textbf {\bibinfo {volume} {4}},\ \bibinfo {pages} {672 }
  (\bibinfo {year} {2021})}\BibitemShut {NoStop}%
\bibitem [{\citenamefont {Yoneda}\ \emph {et~al.}(2018)\citenamefont {Yoneda},
  \citenamefont {Takeda}, \citenamefont {Otsuka}, \citenamefont {Nakajima},
  \citenamefont {Delbecq}, \citenamefont {Allison}, \citenamefont {Honda},
  \citenamefont {Kodera}, \citenamefont {Oda}, \citenamefont {Hoshi},
  \citenamefont {Usami}, \citenamefont {Itoh},\ and\ \citenamefont
  {Tarucha}}]{Yoneda_2018_NatNano_COherence99}%
  \BibitemOpen
  \bibfield  {author} {\bibinfo {author} {\bibfnamefont {J.}~\bibnamefont
  {Yoneda}}, \bibinfo {author} {\bibfnamefont {K.}~\bibnamefont {Takeda}},
  \bibinfo {author} {\bibfnamefont {T.}~\bibnamefont {Otsuka}}, \bibinfo
  {author} {\bibfnamefont {T.}~\bibnamefont {Nakajima}}, \bibinfo {author}
  {\bibfnamefont {M.~R.}\ \bibnamefont {Delbecq}}, \bibinfo {author}
  {\bibfnamefont {G.}~\bibnamefont {Allison}}, \bibinfo {author} {\bibfnamefont
  {T.}~\bibnamefont {Honda}}, \bibinfo {author} {\bibfnamefont
  {T.}~\bibnamefont {Kodera}}, \bibinfo {author} {\bibfnamefont
  {S.}~\bibnamefont {Oda}}, \bibinfo {author} {\bibfnamefont {Y.}~\bibnamefont
  {Hoshi}}, \bibinfo {author} {\bibfnamefont {N.}~\bibnamefont {Usami}},
  \bibinfo {author} {\bibfnamefont {K.~M.}\ \bibnamefont {Itoh}}, \ and\
  \bibinfo {author} {\bibfnamefont {S.}~\bibnamefont {Tarucha}},\ }\href
  {\doibase 10.1038/s41565-017-0014-x} {\bibfield  {journal} {\bibinfo
  {journal} {Nature Nanotechnology}\ }\textbf {\bibinfo {volume} {13}},\
  \bibinfo {pages} {102} (\bibinfo {year} {2018})}\BibitemShut {NoStop}%
\bibitem [{\citenamefont {Lawrie}\ \emph {et~al.}(2021)\citenamefont {Lawrie},
  \citenamefont {Russ}, \citenamefont {van Riggelen}, \citenamefont
  {Hendrickx}, \citenamefont {de~Snoo}, \citenamefont {Sammak}, \citenamefont
  {Scappucci},\ and\ \citenamefont
  {Veldhorst}}]{Lawrie_2021_arxiv_SimultaneousDriving}%
  \BibitemOpen
  \bibfield  {author} {\bibinfo {author} {\bibfnamefont {W.~I.~L.}\
  \bibnamefont {Lawrie}}, \bibinfo {author} {\bibfnamefont {M.}~\bibnamefont
  {Russ}}, \bibinfo {author} {\bibfnamefont {F.}~\bibnamefont {van Riggelen}},
  \bibinfo {author} {\bibfnamefont {N.~W.}\ \bibnamefont {Hendrickx}}, \bibinfo
  {author} {\bibfnamefont {S.~L.}\ \bibnamefont {de~Snoo}}, \bibinfo {author}
  {\bibfnamefont {A.}~\bibnamefont {Sammak}}, \bibinfo {author} {\bibfnamefont
  {G.}~\bibnamefont {Scappucci}}, \ and\ \bibinfo {author} {\bibfnamefont
  {M.}~\bibnamefont {Veldhorst}},\ }\href@noop {} {\enquote {\bibinfo {title}
  {Simultaneous driving of semiconductor spin qubits at the fault-tolerant
  threshold},}\ } (\bibinfo {year} {2021}),\ \Eprint
  {http://arxiv.org/abs/2109.07837} {arXiv:2109.07837 [cond-mat.mes-hall]}
  \BibitemShut {NoStop}%
\bibitem [{\citenamefont {Xue}\ \emph {et~al.}(2022)\citenamefont {Xue},
  \citenamefont {Russ}, \citenamefont {Samkharadze}, \citenamefont {Undseth},
  \citenamefont {Sammak}, \citenamefont {Scappucci},\ and\ \citenamefont
  {Vandersypen}}]{Xue2022}%
  \BibitemOpen
  \bibfield  {author} {\bibinfo {author} {\bibfnamefont {X.}~\bibnamefont
  {Xue}}, \bibinfo {author} {\bibfnamefont {M.}~\bibnamefont {Russ}}, \bibinfo
  {author} {\bibfnamefont {N.}~\bibnamefont {Samkharadze}}, \bibinfo {author}
  {\bibfnamefont {B.}~\bibnamefont {Undseth}}, \bibinfo {author} {\bibfnamefont
  {A.}~\bibnamefont {Sammak}}, \bibinfo {author} {\bibfnamefont
  {G.}~\bibnamefont {Scappucci}}, \ and\ \bibinfo {author} {\bibfnamefont
  {L.~M.~K.}\ \bibnamefont {Vandersypen}},\ }\href {\doibase
  10.1038/s41586-021-04273-w} {\bibfield  {journal} {\bibinfo  {journal}
  {Nature}\ }\textbf {\bibinfo {volume} {601}},\ \bibinfo {pages} {343}
  (\bibinfo {year} {2022})}\BibitemShut {NoStop}%
\bibitem [{\citenamefont {Noiri}\ \emph {et~al.}(2022)\citenamefont {Noiri},
  \citenamefont {Takeda}, \citenamefont {Nakajima}, \citenamefont {Kobayashi},
  \citenamefont {Sammak}, \citenamefont {Scappucci},\ and\ \citenamefont
  {Tarucha}}]{Noiri_2022_Nature_ErrorTheshold}%
  \BibitemOpen
  \bibfield  {author} {\bibinfo {author} {\bibfnamefont {A.}~\bibnamefont
  {Noiri}}, \bibinfo {author} {\bibfnamefont {K.}~\bibnamefont {Takeda}},
  \bibinfo {author} {\bibfnamefont {T.}~\bibnamefont {Nakajima}}, \bibinfo
  {author} {\bibfnamefont {T.}~\bibnamefont {Kobayashi}}, \bibinfo {author}
  {\bibfnamefont {A.}~\bibnamefont {Sammak}}, \bibinfo {author} {\bibfnamefont
  {G.}~\bibnamefont {Scappucci}}, \ and\ \bibinfo {author} {\bibfnamefont
  {S.}~\bibnamefont {Tarucha}},\ }\href {\doibase 10.1038/s41586-021-04182-y}
  {\bibfield  {journal} {\bibinfo  {journal} {Nature}\ }\textbf {\bibinfo
  {volume} {601}},\ \bibinfo {pages} {338} (\bibinfo {year}
  {2022})}\BibitemShut {NoStop}%
\bibitem [{\citenamefont {Takeda}\ \emph {et~al.}(2022)\citenamefont {Takeda},
  \citenamefont {Noiri}, \citenamefont {Nakajima}, \citenamefont {Kobayashi},\
  and\ \citenamefont {Tarucha}}]{Takeda2022}%
  \BibitemOpen
  \bibfield  {author} {\bibinfo {author} {\bibfnamefont {K.}~\bibnamefont
  {Takeda}}, \bibinfo {author} {\bibfnamefont {A.}~\bibnamefont {Noiri}},
  \bibinfo {author} {\bibfnamefont {T.}~\bibnamefont {Nakajima}}, \bibinfo
  {author} {\bibfnamefont {T.}~\bibnamefont {Kobayashi}}, \ and\ \bibinfo
  {author} {\bibfnamefont {S.}~\bibnamefont {Tarucha}},\ }\href {\doibase
  10.1038/s41586-022-04986-6} {\bibfield  {journal} {\bibinfo  {journal}
  {Nature}\ }\textbf {\bibinfo {volume} {608}},\ \bibinfo {pages} {682}
  (\bibinfo {year} {2022})}\BibitemShut {NoStop}%
\bibitem [{\citenamefont {van Riggelen}\ \emph {et~al.}(2022)\citenamefont {van
  Riggelen}, \citenamefont {Lawrie}, \citenamefont {Russ}, \citenamefont
  {Hendrickx}, \citenamefont {Sammak}, \citenamefont {Rispler}, \citenamefont
  {Terhal}, \citenamefont {Scappucci},\ and\ \citenamefont
  {Veldhorst}}]{VanRiggelenPhase}%
  \BibitemOpen
  \bibfield  {author} {\bibinfo {author} {\bibfnamefont {F.}~\bibnamefont {van
  Riggelen}}, \bibinfo {author} {\bibfnamefont {W.~I.~L.}\ \bibnamefont
  {Lawrie}}, \bibinfo {author} {\bibfnamefont {M.}~\bibnamefont {Russ}},
  \bibinfo {author} {\bibfnamefont {N.~W.}\ \bibnamefont {Hendrickx}}, \bibinfo
  {author} {\bibfnamefont {A.}~\bibnamefont {Sammak}}, \bibinfo {author}
  {\bibfnamefont {M.}~\bibnamefont {Rispler}}, \bibinfo {author} {\bibfnamefont
  {B.~M.}\ \bibnamefont {Terhal}}, \bibinfo {author} {\bibfnamefont
  {G.}~\bibnamefont {Scappucci}}, \ and\ \bibinfo {author} {\bibfnamefont
  {M.}~\bibnamefont {Veldhorst}},\ }\href {\doibase 10.1038/s41534-022-00639-8}
  {\bibfield  {journal} {\bibinfo  {journal} {npj Quantum Information}\
  }\textbf {\bibinfo {volume} {8}},\ \bibinfo {pages} {124} (\bibinfo {year}
  {2022})}\BibitemShut {NoStop}%
\bibitem [{\citenamefont {Borsoi}\ \emph {et~al.}(2022)\citenamefont {Borsoi},
  \citenamefont {Hendrickx}, \citenamefont {John}, \citenamefont {Motz},
  \citenamefont {van Riggelen}, \citenamefont {Sammak}, \citenamefont
  {de~Snoo}, \citenamefont {Scappucci},\ and\ \citenamefont
  {Veldhorst}}]{Borsoi2022}%
  \BibitemOpen
  \bibfield  {author} {\bibinfo {author} {\bibfnamefont {F.}~\bibnamefont
  {Borsoi}}, \bibinfo {author} {\bibfnamefont {N.~W.}\ \bibnamefont
  {Hendrickx}}, \bibinfo {author} {\bibfnamefont {V.}~\bibnamefont {John}},
  \bibinfo {author} {\bibfnamefont {S.}~\bibnamefont {Motz}}, \bibinfo {author}
  {\bibfnamefont {F.}~\bibnamefont {van Riggelen}}, \bibinfo {author}
  {\bibfnamefont {A.}~\bibnamefont {Sammak}}, \bibinfo {author} {\bibfnamefont
  {S.~L.}\ \bibnamefont {de~Snoo}}, \bibinfo {author} {\bibfnamefont
  {G.}~\bibnamefont {Scappucci}}, \ and\ \bibinfo {author} {\bibfnamefont
  {M.}~\bibnamefont {Veldhorst}},\ }\href@noop {} {\bibfield  {journal}
  {\bibinfo  {journal} {arXiv}\ } (\bibinfo {year} {2022})},\ \Eprint
  {http://arxiv.org/abs/2209.06609} {arXiv:2209.06609 [cond-mat]} \BibitemShut
  {NoStop}%
\bibitem [{\citenamefont {Vandersypen}\ \emph {et~al.}(2017)\citenamefont
  {Vandersypen}, \citenamefont {Bluhm}, \citenamefont {Clarke}, \citenamefont
  {Dzurak}, \citenamefont {Ishihara}, \citenamefont {Morello}, \citenamefont
  {Reilly}, \citenamefont {Schreiber},\ and\ \citenamefont
  {Veldhorst}}]{Vandersypen_2017_nature_HotDenseCoherent}%
  \BibitemOpen
  \bibfield  {author} {\bibinfo {author} {\bibfnamefont {L.~M.~K.}\
  \bibnamefont {Vandersypen}}, \bibinfo {author} {\bibfnamefont
  {H.}~\bibnamefont {Bluhm}}, \bibinfo {author} {\bibfnamefont {J.~S.}\
  \bibnamefont {Clarke}}, \bibinfo {author} {\bibfnamefont {A.~S.}\
  \bibnamefont {Dzurak}}, \bibinfo {author} {\bibfnamefont {R.}~\bibnamefont
  {Ishihara}}, \bibinfo {author} {\bibfnamefont {A.}~\bibnamefont {Morello}},
  \bibinfo {author} {\bibfnamefont {D.~J.}\ \bibnamefont {Reilly}}, \bibinfo
  {author} {\bibfnamefont {L.~R.}\ \bibnamefont {Schreiber}}, \ and\ \bibinfo
  {author} {\bibfnamefont {M.}~\bibnamefont {Veldhorst}},\ }\href {\doibase
  10.1038/s41534-017-0038-y} {\bibfield  {journal} {\bibinfo  {journal} {npj
  Quantum Information}\ }\textbf {\bibinfo {volume} {3}},\ \bibinfo {pages}
  {34} (\bibinfo {year} {2017})}\BibitemShut {NoStop}%
\bibitem [{\citenamefont {Wang}\ \emph
  {et~al.}(2023{\natexlab{a}})\citenamefont {Wang}, \citenamefont {Feng},
  \citenamefont {Serrano}, \citenamefont {Gilbert}, \citenamefont {Leon},
  \citenamefont {Tanttu}, \citenamefont {Mai}, \citenamefont {Liang},
  \citenamefont {Huang}, \citenamefont {Su}, \citenamefont {Lim}, \citenamefont
  {Hudson}, \citenamefont {Escott}, \citenamefont {Morello}, \citenamefont
  {Yang}, \citenamefont {Dzurak}, \citenamefont {Saraiva},\ and\ \citenamefont
  {Laucht}}]{Wang_2023_AdvancedMat_Jellybean}%
  \BibitemOpen
  \bibfield  {author} {\bibinfo {author} {\bibfnamefont {Z.}~\bibnamefont
  {Wang}}, \bibinfo {author} {\bibfnamefont {M.}~\bibnamefont {Feng}}, \bibinfo
  {author} {\bibfnamefont {S.}~\bibnamefont {Serrano}}, \bibinfo {author}
  {\bibfnamefont {W.}~\bibnamefont {Gilbert}}, \bibinfo {author} {\bibfnamefont
  {R.~C.~C.}\ \bibnamefont {Leon}}, \bibinfo {author} {\bibfnamefont
  {T.}~\bibnamefont {Tanttu}}, \bibinfo {author} {\bibfnamefont
  {P.}~\bibnamefont {Mai}}, \bibinfo {author} {\bibfnamefont {D.}~\bibnamefont
  {Liang}}, \bibinfo {author} {\bibfnamefont {J.~Y.}\ \bibnamefont {Huang}},
  \bibinfo {author} {\bibfnamefont {Y.}~\bibnamefont {Su}}, \bibinfo {author}
  {\bibfnamefont {W.~H.}\ \bibnamefont {Lim}}, \bibinfo {author} {\bibfnamefont
  {F.~E.}\ \bibnamefont {Hudson}}, \bibinfo {author} {\bibfnamefont {C.~C.}\
  \bibnamefont {Escott}}, \bibinfo {author} {\bibfnamefont {A.}~\bibnamefont
  {Morello}}, \bibinfo {author} {\bibfnamefont {C.~H.}\ \bibnamefont {Yang}},
  \bibinfo {author} {\bibfnamefont {A.~S.}\ \bibnamefont {Dzurak}}, \bibinfo
  {author} {\bibfnamefont {A.}~\bibnamefont {Saraiva}}, \ and\ \bibinfo
  {author} {\bibfnamefont {A.}~\bibnamefont {Laucht}},\ }\href {\doibase
  https://doi.org/10.1002/adma.202208557} {\bibfield  {journal} {\bibinfo
  {journal} {Advanced Materials}\ }\textbf {\bibinfo {volume} {n/a}},\ \bibinfo
  {pages} {2208557} (\bibinfo {year} {2023}{\natexlab{a}})}\BibitemShut
  {NoStop}%
\bibitem [{\citenamefont {Borjans}\ \emph {et~al.}(2020)\citenamefont
  {Borjans}, \citenamefont {Croot}, \citenamefont {Mi}, \citenamefont
  {Gullans},\ and\ \citenamefont
  {Petta}}]{Borjans_2020_Nature_PhotonMediatedLink}%
  \BibitemOpen
  \bibfield  {author} {\bibinfo {author} {\bibfnamefont {F.}~\bibnamefont
  {Borjans}}, \bibinfo {author} {\bibfnamefont {X.~G.}\ \bibnamefont {Croot}},
  \bibinfo {author} {\bibfnamefont {X.}~\bibnamefont {Mi}}, \bibinfo {author}
  {\bibfnamefont {M.~J.}\ \bibnamefont {Gullans}}, \ and\ \bibinfo {author}
  {\bibfnamefont {J.~R.}\ \bibnamefont {Petta}},\ }\href {\doibase
  10.1038/s41586-019-1867-y} {\bibfield  {journal} {\bibinfo  {journal}
  {Nature}\ }\textbf {\bibinfo {volume} {577}},\ \bibinfo {pages} {195}
  (\bibinfo {year} {2020})}\BibitemShut {NoStop}%
\bibitem [{\citenamefont {Yoneda}\ \emph {et~al.}(2021)\citenamefont {Yoneda},
  \citenamefont {Huang}, \citenamefont {Feng}, \citenamefont {Yang},
  \citenamefont {Chan}, \citenamefont {Tanttu}, \citenamefont {Gilbert},
  \citenamefont {Leon}, \citenamefont {Hudson}, \citenamefont {Itoh},
  \citenamefont {Morello}, \citenamefont {Bartlett}, \citenamefont {Laucht},
  \citenamefont {Saraiva},\ and\ \citenamefont
  {Dzurak}}]{Yoneda_2021_NatureComm_CoherentTransport}%
  \BibitemOpen
  \bibfield  {author} {\bibinfo {author} {\bibfnamefont {J.}~\bibnamefont
  {Yoneda}}, \bibinfo {author} {\bibfnamefont {W.}~\bibnamefont {Huang}},
  \bibinfo {author} {\bibfnamefont {M.}~\bibnamefont {Feng}}, \bibinfo {author}
  {\bibfnamefont {C.~H.}\ \bibnamefont {Yang}}, \bibinfo {author}
  {\bibfnamefont {K.~W.}\ \bibnamefont {Chan}}, \bibinfo {author}
  {\bibfnamefont {T.}~\bibnamefont {Tanttu}}, \bibinfo {author} {\bibfnamefont
  {W.}~\bibnamefont {Gilbert}}, \bibinfo {author} {\bibfnamefont {R.~C.~C.}\
  \bibnamefont {Leon}}, \bibinfo {author} {\bibfnamefont {F.~E.}\ \bibnamefont
  {Hudson}}, \bibinfo {author} {\bibfnamefont {K.~M.}\ \bibnamefont {Itoh}},
  \bibinfo {author} {\bibfnamefont {A.}~\bibnamefont {Morello}}, \bibinfo
  {author} {\bibfnamefont {S.~D.}\ \bibnamefont {Bartlett}}, \bibinfo {author}
  {\bibfnamefont {A.}~\bibnamefont {Laucht}}, \bibinfo {author} {\bibfnamefont
  {A.}~\bibnamefont {Saraiva}}, \ and\ \bibinfo {author} {\bibfnamefont
  {A.~S.}\ \bibnamefont {Dzurak}},\ }\href {\doibase
  10.1038/s41467-021-24371-7} {\bibfield  {journal} {\bibinfo  {journal}
  {Nature Communications}\ }\textbf {\bibinfo {volume} {12}},\ \bibinfo {pages}
  {4114} (\bibinfo {year} {2021})}\BibitemShut {NoStop}%
\bibitem [{\citenamefont {Tosato}\ \emph {et~al.}(2022)\citenamefont {Tosato},
  \citenamefont {Ferrari}, \citenamefont {Sammak}, \citenamefont {Hamilton},
  \citenamefont {Veldhorst}, \citenamefont {Virgilio},\ and\ \citenamefont
  {Scappucci}}]{Tosato2022}%
  \BibitemOpen
  \bibfield  {author} {\bibinfo {author} {\bibfnamefont {A.}~\bibnamefont
  {Tosato}}, \bibinfo {author} {\bibfnamefont {B.}~\bibnamefont {Ferrari}},
  \bibinfo {author} {\bibfnamefont {A.}~\bibnamefont {Sammak}}, \bibinfo
  {author} {\bibfnamefont {A.~R.}\ \bibnamefont {Hamilton}}, \bibinfo {author}
  {\bibfnamefont {M.}~\bibnamefont {Veldhorst}}, \bibinfo {author}
  {\bibfnamefont {M.}~\bibnamefont {Virgilio}}, \ and\ \bibinfo {author}
  {\bibfnamefont {G.}~\bibnamefont {Scappucci}},\ }\href {\doibase
  10.1002/qute.202100167} {\bibfield  {journal} {\bibinfo  {journal} {Advanced
  Quantum Technologies}\ }\textbf {\bibinfo {volume} {5}},\ \bibinfo {pages}
  {2100167} (\bibinfo {year} {2022})}\BibitemShut {NoStop}%
\bibitem [{\citenamefont {Laroche}\ \emph {et~al.}(2015)\citenamefont
  {Laroche}, \citenamefont {Huang}, \citenamefont {Nielsen}, \citenamefont
  {Liu}, \citenamefont {Li},\ and\ \citenamefont {Lu}}]{Laroche2015}%
  \BibitemOpen
  \bibfield  {author} {\bibinfo {author} {\bibfnamefont {D.}~\bibnamefont
  {Laroche}}, \bibinfo {author} {\bibfnamefont {S.-H.}\ \bibnamefont {Huang}},
  \bibinfo {author} {\bibfnamefont {E.}~\bibnamefont {Nielsen}}, \bibinfo
  {author} {\bibfnamefont {C.~W.}\ \bibnamefont {Liu}}, \bibinfo {author}
  {\bibfnamefont {J.-Y.}\ \bibnamefont {Li}}, \ and\ \bibinfo {author}
  {\bibfnamefont {T.~M.}\ \bibnamefont {Lu}},\ }\href {\doibase
  10.1063/1.4917296} {\bibfield  {journal} {\bibinfo  {journal} {Applied
  Physics Letters}\ }\textbf {\bibinfo {volume} {106}} (\bibinfo {year}
  {2015}),\ 10.1063/1.4917296},\ \bibinfo {note} {143503},\ \Eprint
  {http://arxiv.org/abs/https://pubs.aip.org/aip/apl/article-pdf/doi/10.1063/1.4917296/13050894/143503\_1\_online.pdf}
  {https://pubs.aip.org/aip/apl/article-pdf/doi/10.1063/1.4917296/13050894/143503\_1\_online.pdf}
  \BibitemShut {NoStop}%
\bibitem [{\citenamefont {Fujita}\ \emph {et~al.}(2017)\citenamefont {Fujita},
  \citenamefont {Baart}, \citenamefont {Reichl}, \citenamefont {Wegscheider},\
  and\ \citenamefont {Vandersypen}}]{Fujita_2017_npjQuantum_CoherentShuttle}%
  \BibitemOpen
  \bibfield  {author} {\bibinfo {author} {\bibfnamefont {T.}~\bibnamefont
  {Fujita}}, \bibinfo {author} {\bibfnamefont {T.~A.}\ \bibnamefont {Baart}},
  \bibinfo {author} {\bibfnamefont {C.}~\bibnamefont {Reichl}}, \bibinfo
  {author} {\bibfnamefont {W.}~\bibnamefont {Wegscheider}}, \ and\ \bibinfo
  {author} {\bibfnamefont {L.~M.~K.}\ \bibnamefont {Vandersypen}},\ }\href
  {\doibase 10.1038/s41534-017-0024-4} {\bibfield  {journal} {\bibinfo
  {journal} {npj Quantum Information}\ }\textbf {\bibinfo {volume} {3}},\
  \bibinfo {pages} {22} (\bibinfo {year} {2017})}\BibitemShut {NoStop}%
\bibitem [{\citenamefont {Mutter}\ and\ \citenamefont
  {Burkard}(2021)}]{Mutter_2021_PRRES_FloppingMode}%
  \BibitemOpen
  \bibfield  {author} {\bibinfo {author} {\bibfnamefont {P.~M.}\ \bibnamefont
  {Mutter}}\ and\ \bibinfo {author} {\bibfnamefont {G.}~\bibnamefont
  {Burkard}},\ }\href {\doibase 10.1103/PhysRevResearch.3.013194} {\bibfield
  {journal} {\bibinfo  {journal} {Physical Review Research}\ }\textbf {\bibinfo
  {volume} {3}},\ \bibinfo {pages} {13194} (\bibinfo {year}
  {2021})}\BibitemShut {NoStop}%
\bibitem [{\citenamefont {Malissa}\ \emph {et~al.}(2006)\citenamefont
  {Malissa}, \citenamefont {Gruber}, \citenamefont {Pachinger}, \citenamefont
  {Schäffler}, \citenamefont {Jantsch},\ and\ \citenamefont
  {Wilamowski}}]{Malissa2006}%
  \BibitemOpen
  \bibfield  {author} {\bibinfo {author} {\bibfnamefont {H.}~\bibnamefont
  {Malissa}}, \bibinfo {author} {\bibfnamefont {D.}~\bibnamefont {Gruber}},
  \bibinfo {author} {\bibfnamefont {D.}~\bibnamefont {Pachinger}}, \bibinfo
  {author} {\bibfnamefont {F.}~\bibnamefont {Schäffler}}, \bibinfo {author}
  {\bibfnamefont {W.}~\bibnamefont {Jantsch}}, \ and\ \bibinfo {author}
  {\bibfnamefont {Z.}~\bibnamefont {Wilamowski}},\ }\href {\doibase
  10.1016/j.spmi.2005.09.001} {\bibfield  {journal} {\bibinfo  {journal}
  {Superlattices and Microstructures}\ }\textbf {\bibinfo {volume} {39}},\
  \bibinfo {pages} {414} (\bibinfo {year} {2006})}\BibitemShut {NoStop}%
\bibitem [{\citenamefont {Scappucci}\ \emph {et~al.}(2020)\citenamefont
  {Scappucci}, \citenamefont {Kloeffel}, \citenamefont {Zwanenburg},
  \citenamefont {Loss}, \citenamefont {Myronov}, \citenamefont {Zhang},
  \citenamefont {De~Franceschi}, \citenamefont {Katsaros},\ and\ \citenamefont
  {Veldhorst}}]{Scappucci2020TheRoute}%
  \BibitemOpen
  \bibfield  {author} {\bibinfo {author} {\bibfnamefont {G.}~\bibnamefont
  {Scappucci}}, \bibinfo {author} {\bibfnamefont {C.}~\bibnamefont {Kloeffel}},
  \bibinfo {author} {\bibfnamefont {F.~A.}\ \bibnamefont {Zwanenburg}},
  \bibinfo {author} {\bibfnamefont {D.}~\bibnamefont {Loss}}, \bibinfo {author}
  {\bibfnamefont {M.}~\bibnamefont {Myronov}}, \bibinfo {author} {\bibfnamefont
  {J.~J.}\ \bibnamefont {Zhang}}, \bibinfo {author} {\bibfnamefont
  {S.}~\bibnamefont {De~Franceschi}}, \bibinfo {author} {\bibfnamefont
  {G.}~\bibnamefont {Katsaros}}, \ and\ \bibinfo {author} {\bibfnamefont
  {M.}~\bibnamefont {Veldhorst}},\ }\href {\doibase 10.1038/s41578-020-00262-z}
  {\bibfield  {journal} {\bibinfo  {journal} {Nature Reviews Materials}\
  }\textbf {\bibinfo {volume} {6}},\ \bibinfo {pages} {926–943} (\bibinfo
  {year} {2020})}\BibitemShut {NoStop}%
\bibitem [{\citenamefont {Wang}\ \emph
  {et~al.}(2023{\natexlab{b}})\citenamefont {Wang}, \citenamefont {Déprez},
  \citenamefont {Tidjani}, \citenamefont {Lawrie}, \citenamefont {Hendrickx},
  \citenamefont {Sammak}, \citenamefont {Scappucci},\ and\ \citenamefont
  {Veldhorst}}]{Wang2022_RVB}%
  \BibitemOpen
  \bibfield  {author} {\bibinfo {author} {\bibfnamefont {C.-A.}\ \bibnamefont
  {Wang}}, \bibinfo {author} {\bibfnamefont {C.}~\bibnamefont {Déprez}},
  \bibinfo {author} {\bibfnamefont {H.}~\bibnamefont {Tidjani}}, \bibinfo
  {author} {\bibfnamefont {W.~I.~L.}\ \bibnamefont {Lawrie}}, \bibinfo {author}
  {\bibfnamefont {N.~W.}\ \bibnamefont {Hendrickx}}, \bibinfo {author}
  {\bibfnamefont {A.}~\bibnamefont {Sammak}}, \bibinfo {author} {\bibfnamefont
  {G.}~\bibnamefont {Scappucci}}, \ and\ \bibinfo {author} {\bibfnamefont
  {M.}~\bibnamefont {Veldhorst}},\ }\href@noop {} {\bibfield  {journal}
  {\bibinfo  {journal} {NPJ Quantum Information}\ } (\bibinfo {year}
  {2023}{\natexlab{b}})}\BibitemShut {NoStop}%
\bibitem [{\citenamefont {Conti}\ \emph {et~al.}(2021)\citenamefont {Conti},
  \citenamefont {Saberi-Pouya}, \citenamefont {Perali}, \citenamefont
  {Virgilio}, \citenamefont {Peeters}, \citenamefont {Hamilton}, \citenamefont
  {Scappucci},\ and\ \citenamefont
  {Neilson}}]{Conti_2021_npjQuantum_BilayerBEC}%
  \BibitemOpen
  \bibfield  {author} {\bibinfo {author} {\bibfnamefont {S.}~\bibnamefont
  {Conti}}, \bibinfo {author} {\bibfnamefont {S.}~\bibnamefont {Saberi-Pouya}},
  \bibinfo {author} {\bibfnamefont {A.}~\bibnamefont {Perali}}, \bibinfo
  {author} {\bibfnamefont {M.}~\bibnamefont {Virgilio}}, \bibinfo {author}
  {\bibfnamefont {F.~M.}\ \bibnamefont {Peeters}}, \bibinfo {author}
  {\bibfnamefont {A.~R.}\ \bibnamefont {Hamilton}}, \bibinfo {author}
  {\bibfnamefont {G.}~\bibnamefont {Scappucci}}, \ and\ \bibinfo {author}
  {\bibfnamefont {D.}~\bibnamefont {Neilson}},\ }\href {\doibase
  10.1038/s41535-021-00344-3} {\bibfield  {journal} {\bibinfo  {journal} {npj
  Quantum Materials}\ }\textbf {\bibinfo {volume} {6}},\ \bibinfo {pages} {41}
  (\bibinfo {year} {2021})}\BibitemShut {NoStop}%
\bibitem [{\citenamefont {Nandi}\ \emph {et~al.}(2012)\citenamefont {Nandi},
  \citenamefont {Finck}, \citenamefont {Eisenstein}, \citenamefont {Pfeiffer},\
  and\ \citenamefont {West}}]{Nandi2012}%
  \BibitemOpen
  \bibfield  {author} {\bibinfo {author} {\bibfnamefont {D.}~\bibnamefont
  {Nandi}}, \bibinfo {author} {\bibfnamefont {A.~D.~K.}\ \bibnamefont {Finck}},
  \bibinfo {author} {\bibfnamefont {J.~P.}\ \bibnamefont {Eisenstein}},
  \bibinfo {author} {\bibfnamefont {L.~N.}\ \bibnamefont {Pfeiffer}}, \ and\
  \bibinfo {author} {\bibfnamefont {K.~W.}\ \bibnamefont {West}},\ }\href
  {\doibase 10.1038/nature11302} {\bibfield  {journal} {\bibinfo  {journal}
  {Nature}\ }\textbf {\bibinfo {volume} {488}},\ \bibinfo {pages} {481}
  (\bibinfo {year} {2012})}\BibitemShut {NoStop}%
\bibitem [{\citenamefont {Hamo}\ \emph {et~al.}(2016)\citenamefont {Hamo},
  \citenamefont {Benyamini}, \citenamefont {Shapir}, \citenamefont {Khivrich},
  \citenamefont {Waissman}, \citenamefont {Kaasbjerg}, \citenamefont {Oreg},
  \citenamefont {von Oppen},\ and\ \citenamefont {Ilani}}]{Hamo2016}%
  \BibitemOpen
  \bibfield  {author} {\bibinfo {author} {\bibfnamefont {A.}~\bibnamefont
  {Hamo}}, \bibinfo {author} {\bibfnamefont {A.}~\bibnamefont {Benyamini}},
  \bibinfo {author} {\bibfnamefont {I.}~\bibnamefont {Shapir}}, \bibinfo
  {author} {\bibfnamefont {I.}~\bibnamefont {Khivrich}}, \bibinfo {author}
  {\bibfnamefont {J.}~\bibnamefont {Waissman}}, \bibinfo {author}
  {\bibfnamefont {K.}~\bibnamefont {Kaasbjerg}}, \bibinfo {author}
  {\bibfnamefont {Y.}~\bibnamefont {Oreg}}, \bibinfo {author} {\bibfnamefont
  {F.}~\bibnamefont {von Oppen}}, \ and\ \bibinfo {author} {\bibfnamefont
  {S.}~\bibnamefont {Ilani}},\ }\href {\doibase 10.1038/nature18639} {\bibfield
   {journal} {\bibinfo  {journal} {Nature}\ }\textbf {\bibinfo {volume}
  {535}},\ \bibinfo {pages} {395} (\bibinfo {year} {2016})}\BibitemShut
  {NoStop}%
\bibitem [{\citenamefont {Hong}\ \emph {et~al.}(2018)\citenamefont {Hong},
  \citenamefont {Yoo}, \citenamefont {Park}, \citenamefont {Cho}, \citenamefont
  {Chung}, \citenamefont {Sim}, \citenamefont {Kim}, \citenamefont {Choi},
  \citenamefont {Umansky},\ and\ \citenamefont
  {Mahalu}}]{Hong_2018_PRB_AttractiveCoulomb}%
  \BibitemOpen
  \bibfield  {author} {\bibinfo {author} {\bibfnamefont {C.}~\bibnamefont
  {Hong}}, \bibinfo {author} {\bibfnamefont {G.}~\bibnamefont {Yoo}}, \bibinfo
  {author} {\bibfnamefont {J.}~\bibnamefont {Park}}, \bibinfo {author}
  {\bibfnamefont {M.-K.}\ \bibnamefont {Cho}}, \bibinfo {author} {\bibfnamefont
  {Y.}~\bibnamefont {Chung}}, \bibinfo {author} {\bibfnamefont {H.-S.}\
  \bibnamefont {Sim}}, \bibinfo {author} {\bibfnamefont {D.}~\bibnamefont
  {Kim}}, \bibinfo {author} {\bibfnamefont {H.}~\bibnamefont {Choi}}, \bibinfo
  {author} {\bibfnamefont {V.}~\bibnamefont {Umansky}}, \ and\ \bibinfo
  {author} {\bibfnamefont {D.}~\bibnamefont {Mahalu}},\ }\href {\doibase
  10.1103/PhysRevB.97.241115} {\bibfield  {journal} {\bibinfo  {journal} {Phys.
  Rev. B}\ }\textbf {\bibinfo {volume} {97}},\ \bibinfo {pages} {241115}
  (\bibinfo {year} {2018})}\BibitemShut {NoStop}%
\bibitem [{\citenamefont {Hamilton}\ \emph {et~al.}(1995)\citenamefont
  {Hamilton}, \citenamefont {Linfield}, \citenamefont {Kelly}, \citenamefont
  {Ritchie}, \citenamefont {Jones},\ and\ \citenamefont
  {Pepper}}]{Hamilton1995TransitionHeterostructures}%
  \BibitemOpen
  \bibfield  {author} {\bibinfo {author} {\bibfnamefont {A.~R.}\ \bibnamefont
  {Hamilton}}, \bibinfo {author} {\bibfnamefont {E.~H.}\ \bibnamefont
  {Linfield}}, \bibinfo {author} {\bibfnamefont {M.~J.}\ \bibnamefont {Kelly}},
  \bibinfo {author} {\bibfnamefont {D.~A.}\ \bibnamefont {Ritchie}}, \bibinfo
  {author} {\bibfnamefont {G.~A.}\ \bibnamefont {Jones}}, \ and\ \bibinfo
  {author} {\bibfnamefont {M.}~\bibnamefont {Pepper}},\ }\href {\doibase
  10.1103/PhysRevB.51.17600} {\bibfield  {journal} {\bibinfo  {journal}
  {Physical Review B}\ }\textbf {\bibinfo {volume} {51}},\ \bibinfo {pages}
  {17600} (\bibinfo {year} {1995})}\BibitemShut {NoStop}%
\bibitem [{\citenamefont {van~der Wiel}\ \emph {et~al.}(2002)\citenamefont
  {van~der Wiel}, \citenamefont {De~Franceschi}, \citenamefont {Elzerman},
  \citenamefont {Fujisawa}, \citenamefont {Tarucha},\ and\ \citenamefont
  {Kouwenhoven}}]{VanDerWiel2003}%
  \BibitemOpen
  \bibfield  {author} {\bibinfo {author} {\bibfnamefont {W.~G.}\ \bibnamefont
  {van~der Wiel}}, \bibinfo {author} {\bibfnamefont {S.}~\bibnamefont
  {De~Franceschi}}, \bibinfo {author} {\bibfnamefont {J.~M.}\ \bibnamefont
  {Elzerman}}, \bibinfo {author} {\bibfnamefont {T.}~\bibnamefont {Fujisawa}},
  \bibinfo {author} {\bibfnamefont {S.}~\bibnamefont {Tarucha}}, \ and\
  \bibinfo {author} {\bibfnamefont {L.~P.}\ \bibnamefont {Kouwenhoven}},\
  }\href {\doibase 10.1103/RevModPhys.75.1} {\bibfield  {journal} {\bibinfo
  {journal} {Rev. Mod. Phys.}\ }\textbf {\bibinfo {volume} {75}},\ \bibinfo
  {pages} {1} (\bibinfo {year} {2002})}\BibitemShut {NoStop}%
\bibitem [{Ans(2022)}]{Ansys}%
  \BibitemOpen
  \href@noop {} {\enquote {\bibinfo {title} {Ansys® {Electronics} {Desktop},
  {Release} 22.2},}\ } (\bibinfo {year} {2022})\BibitemShut {NoStop}%
\bibitem [{\citenamefont {Birner}\ \emph {et~al.}(2007)\citenamefont {Birner},
  \citenamefont {Zibold}, \citenamefont {Andlauer}, \citenamefont {Kubis},
  \citenamefont {Sabathil}, \citenamefont {Trellakis},\ and\ \citenamefont
  {Vogl}}]{nextnano}%
  \BibitemOpen
  \bibfield  {author} {\bibinfo {author} {\bibfnamefont {S.}~\bibnamefont
  {Birner}}, \bibinfo {author} {\bibfnamefont {T.}~\bibnamefont {Zibold}},
  \bibinfo {author} {\bibfnamefont {T.}~\bibnamefont {Andlauer}}, \bibinfo
  {author} {\bibfnamefont {T.}~\bibnamefont {Kubis}}, \bibinfo {author}
  {\bibfnamefont {M.}~\bibnamefont {Sabathil}}, \bibinfo {author}
  {\bibfnamefont {A.}~\bibnamefont {Trellakis}}, \ and\ \bibinfo {author}
  {\bibfnamefont {P.}~\bibnamefont {Vogl}},\ }\href {\doibase
  10.1109/TED.2007.902871} {\bibfield  {journal} {\bibinfo  {journal} {IEEE
  Transactions on Electron Devices}\ }\textbf {\bibinfo {volume} {54}},\
  \bibinfo {pages} {2137} (\bibinfo {year} {2007})}\BibitemShut {NoStop}%
\bibitem [{\citenamefont {Groner}\ \emph {et~al.}(2002)\citenamefont {Groner},
  \citenamefont {Elam}, \citenamefont {Fabreguette},\ and\ \citenamefont
  {George}}]{Groner_ThinSolidFilms_2002_AlOxProperties}%
  \BibitemOpen
  \bibfield  {author} {\bibinfo {author} {\bibfnamefont {M.}~\bibnamefont
  {Groner}}, \bibinfo {author} {\bibfnamefont {J.}~\bibnamefont {Elam}},
  \bibinfo {author} {\bibfnamefont {F.}~\bibnamefont {Fabreguette}}, \ and\
  \bibinfo {author} {\bibfnamefont {S.}~\bibnamefont {George}},\ }\href
  {\doibase https://doi.org/10.1016/S0040-6090(02)00438-8} {\bibfield
  {journal} {\bibinfo  {journal} {Thin Solid Films}\ }\textbf {\bibinfo
  {volume} {413}},\ \bibinfo {pages} {186} (\bibinfo {year}
  {2002})}\BibitemShut {NoStop}%
\bibitem [{\citenamefont {Dunlap}\ and\ \citenamefont
  {Watters}(1953)}]{Dunlap_PhysRev_1953_SiGeDielectricConstants}%
  \BibitemOpen
  \bibfield  {author} {\bibinfo {author} {\bibfnamefont {W.~C.}\ \bibnamefont
  {Dunlap}}\ and\ \bibinfo {author} {\bibfnamefont {R.~L.}\ \bibnamefont
  {Watters}},\ }\href {\doibase 10.1103/PhysRev.92.1396} {\bibfield  {journal}
  {\bibinfo  {journal} {Phys. Rev.}\ }\textbf {\bibinfo {volume} {92}},\
  \bibinfo {pages} {1396} (\bibinfo {year} {1953})}\BibitemShut {NoStop}%
\bibitem [{\citenamefont {Sze}(1981)}]{Sze_1981_PropertiesSiO2}%
  \BibitemOpen
  \bibfield  {author} {\bibinfo {author} {\bibfnamefont {S.}~\bibnamefont
  {Sze}}\ }(\bibinfo  {publisher} {New York: Wiley},\ \bibinfo {year}
  {1981})\BibitemShut {NoStop}%
\end{thebibliography}%
\end{document}